%% file: main.tex
\documentclass[sigconf,10pt]{acmart}

\usepackage[english]{babel}
\usepackage{blindtext}
\renewcommand\footnotetextcopyrightpermission[1]{} % removes footnote with conference info
\setcopyright{none}

\usepackage{amsmath,amsfonts}

\usepackage{graphicx}
\usepackage{appendix}
\usepackage{hyperref}
\usepackage{outlines}
\usepackage{pgfplots}
\usepackage{xcolor}
\usepackage{comment}
\usepackage{epigraph}
\usepackage{enumitem}
\usepackage{epigraph}
\usepackage{listings}
\usepackage{caption}
\usepackage{subcaption}
\usepackage{tikz}
\usepackage{xspace}
\usepackage{multirow}

\usepackage{listings}
\usepackage{algorithm}
\usepackage{algpseudocode}
\usetikzlibrary{arrows,shapes,positioning,fit,decorations.pathreplacing, shapes.geometric}
\usepackage{pifont}
\usepackage{enumitem}
\setitemize{nosep, leftmargin=1.5em}
\usepackage{caption}
\algnewcommand{\algorithmicand}{\textbf{ and }}
\algnewcommand{\algorithmicor}{\textbf{ or }}
\algnewcommand{\OR}{\algorithmicor}
\algnewcommand{\AAND}{\algorithmicand}
\newcommand{\cmark}{\ding{51}}%
\newcommand{\xmark}{\ding{55}}%
\hypersetup{pdfstartview=FitH,pdfpagelayout=SinglePage}

\usepackage[compact]{titlesec}         % you need this package
\titlespacing{\subsubsection}{15pt}{2pt}{2pt}% this reduces space between (sub)sections to 0pt, for example

\newenvironment{tightitemize}%
 {\begin{list}{$\bullet$}{%
 		\setlength{\leftmargin}{10pt}
        \setlength{\itemsep}{0pt}%
        \setlength{\parsep}{0pt}%
        \setlength{\topsep}{0pt}%
        \setlength{\parskip}{0pt}%
        }%
 }%
 {\end{list}}

\definecolor{orange-fill}{HTML}{FFE6CC}
\definecolor{orange-line}{HTML}{D79B00}
\definecolor{green-fill}{HTML}{D5E8D4}
\definecolor{green-line}{HTML}{82B366}
\definecolor{red-fill}{HTML}{F8CECC}
\definecolor{red-line}{HTML}{B85450}
\def\srow[#1][#2](#3:#4:#5:#6)[#7][#8]
{
    \node[circle] (#1) at (axis cs:#8,{#2+0.5}) {#1};
    \node[circle] (a) at (axis cs: {#6+0.5},{#2+0.5}) {#7};
    \fill[fill=orange-fill, draw=orange-line](axis cs:0,#2) rectangle (axis cs:#3,{#2+1} );
    \fill[fill=green-fill, draw=green-line](axis cs:#3,#2) rectangle (axis cs:#4,{#2+1});
    \fill[fill=red-fill, draw=red-line](axis cs:#4,#2) rectangle (axis cs:#5,{#2+1});
    \fill[fill=green-fill, draw=green-line](axis cs:#5,#2) rectangle (axis cs:#6,{#2+1});
}
\def\sdrow[#1][#2](#3:#4:#5:#6:#7)[#8][#9]
{
    \node[circle] (#1) at (axis cs:#9,{#2+0.5}) {#1};
    \node[circle] (a) at (axis cs: {#7+0.5},{#2+0.5}) {#8};
    \fill[fill=orange-fill, draw=orange-line](axis cs:0,#2) rectangle (axis cs:#3,{#2+1} );
    \fill[fill=green-fill, draw=green-line](axis cs:#3,#2) rectangle (axis cs:#4,{#2+1});
    \fill[fill=red-fill, draw=red-line](axis cs:#4,#2) rectangle (axis cs:#5,{#2+1});
    \fill[fill=orange-fill, draw=orange-line](axis cs:#5,#2) rectangle (axis cs:#6,{#2+1});
    \fill[fill=green-fill, draw=green-line](axis cs:#6,#2) rectangle (axis cs:#7,{#2+1});
}

\definecolor{keywordcolor}{rgb}{0.7, 0.1, 0.1}   % red
\definecolor{commentcolor}{rgb}{0.4, 0.4, 0.4}   % grey
\definecolor{symbolcolor}{rgb}{0.0, 0.1, 0.6}    % blue
\definecolor{sortcolor}{rgb}{0.1, 0.5, 0.1}      % green
\usepackage{listings}

\newcommand{\note}[2]{\textcolor[rgb]{#1}{[#2]}}
\renewcommand{\note}[2]{}

\newcommand{\framename}{Virelay\xspace}

\newcommand{\algo}{\texttt{Algorithm}\xspace}
\newcommand{\sys}{\texttt{System}\xspace}
\newcommand{\statet}{\texttt{state}\xspace}
\newcommand{\doneCondition}{\texttt{doneCondition}\xspace}
\newcommand{\nextEvent}{\texttt{nextEvent}\xspace}
\newcommand{\funUpdate}{\texttt{update}\xspace}
\newcommand{\objfun}{\texttt{OF}\xspace}
\newcommand{\assign}{\texttt{assign}\xspace}

\title{A Performance Verification Methodology for Resource Allocation Heuristics}
\author{Saksham Goel, Benjamin Mikek$^+$, Jehad Aly$^+$, Venkat Arun, Ahmed Saeed$^+$, Aditya Akella\\
UT Austin, $^+$Georgia Tech\\
}
\begin{abstract}

Performance verification is a nascent but promising tool for understanding the performance and limitations of heuristics under realistic assumptions. Bespoke performance verification tools have already demonstrated their value in settings like congestion control and packet scheduling. In this paper, we aim to emphasize the broad applicability and utility of performance verification. To that end, we highlight the design principles of performance verification. Then, we leverage that understanding to develop a set of easy-to-follow guidelines that are applicable to a wide range of resource allocation heuristics. In particular, we introduce \framename, a framework that enables heuristic designers to express the behavior of their algorithms and their assumptions about the system in an environment that resembles a discrete-event simulator.  We demonstrate the utility and ease-of-use of \framename by applying it to six diverse case studies. We produce bounds on the performance of classical algorithms, work stealing and SRPT scheduling, under practical assumptions. We demonstrate \framename's expressiveness by capturing existing models for congestion control and packet scheduling, and we verify the observation that TCP unfairness can cause some ML training workloads to spontaneously converge to a state of high network utilization. Finally, we use  \framename to identify two bugs in the Linux CFS load balancer.

\end{abstract}
\begin{document}
\maketitle
\pagestyle{plain}

% \va{I'm thinking the tone of this paper should be to describe our experience, rather than argue novelty or sell an artifact. We motivate the need for theoretical understanding of schedulers. Then we say doing this with pen-and-paper is too hard. Thus we use bounded verification and SMT solvers. The rest of the paper is teaching people how to do this, and show them which parts are challenging and how to approach the problem so it is doable}
% \va{I'm trying to use the Scott Shenker method of writing papers collaboratively where we write bad english in bullet points until a few days before the deadline. This allows us to rearrange points and agree on the flow quickly. Then we can convert into nice prose at the end}

\input{01_introduction}

\input{02_motivation}
\input{03_virelay}
\input{04_work_stealing}

\input{05_linux}

\section{Other Case Studies}

\subsection{Single Core SRPT Scheduling}
\label{sec:srpt}
Consider a system that handles a large number of small tasks on a single processor, interspersed with a few longer tasks. The designer is tempted to use the Shortest Remaining Processing Time First (SRPTF) algorithm because it minimizes the average time to completion~\cite{srpt-queue}. Theoretical analyses  produced bounds on the performance and fairness of SRPTF \cite{stress, srpt-analysis}. However, these results assume both preemption and no task blocking. The designer knows that their tasks often block on disk reads and the overhead of preemption is excessive~\cite{shenango, caladan, demikernel}. We show how \framename can help them determine that SRPTF is \emph{not} a good fit in this case. 
%Shortest remaining processing time first (SRPT) is a processor scheduling algorithm which simply runs the task with least remaining time. It is well known that SRPT is optimal with respect to average time to completion \cite{srpt-queue}. Theoretical analyses have also produced bounds on the performance and fairness of SRPT \cite{stress, srpt-analysis}. However, these results assume both preemption and no task blocking. But tasks may block, for example, to perform IO operations. We therefore consider the case where tasks can alternate between running and blocking, employing a  scheduler on a single processor. This context is reflective of many RPC execution settings where threads run to completion and only yield when they make blocking calls~\cite{shenango, caladan, demikernel}.
To expand knowledge about the performance of SRPT to the blocking--allowed context, we use the \framename framework create a system of constraints and compare SRPT's performance to an optimal scheduler on two metrics: 1)  the average completion time of tasks, and 2) the number of tasks that finish by a deadline. The details of the model are in Appendix~\ref{app:srpt}. We focus here on the results.

First, we impose a bound $\alpha$ on the ratio of minimum running times to maximum blocking times. When blocking times are unbounded (\textit{i.e.,} $\alpha = \infty$), we find that SRPT can perform much worse than the optimal scheduler on both metrics. For average time to completion, our model produces task sets for which SRPT schedules are $N_T-\epsilon$ times worse than the optimal schedule, while for the deadline query, \framename finds that SRPT can perform arbitrarily worse than optimal. This behavior occurs because an ideal scheduler can ``look ahead'' and more efficiently overlay (in time) the blocking periods of multiple tasks to improve performance; concrete example task sets are given in the appendix.

For bounded $\alpha$, whenever $\alpha > N_T-2$, it is possible for the SRPT schedule to have arbitrarily high average completion time (up to the $N_T-\epsilon$ bound). For $\alpha \leq N_T-2$, the worst case average completion time grows with the number of tasks, independent of $\alpha$. For the deadline query, we find a {\em linear} relationship between $\alpha$ and the worst case performance of SRPT. In particular, the optimal schedule can finish $\alpha$ times more tasks than SRPT. Our model is able to confirm this result for values of $N_T$ up to $7$ and values of $\alpha \in [0,7]$. This trend is expected -- as the maximum blocking time increases, more and more running time periods of other tasks can overlap with the blocking period of another, which the ideal scheduler can exploit.

%\vspace{-0.12in}
\begin{comment}
\begin{center}
\noindent\centering\fbox{%
    \parbox{.9\linewidth}{
        If tasks can block, SRPT can perform arbitrarily badly. If blocking time is bounded, there is a linear relationship between maximum blocking time and the number of additional tasks an ideal scheduler could finish compared to SRPT.
    }}
\end{center}
\end{comment}

\subsection{TCP Synchronization in Ring Allreduce Training}
\label{sec:rr}

% \begin{figure}
%     \centering
% \begin{tikzpicture}
% [server/.style={circle,draw=blue!50,fill=blue!20,thick,inner sep=0pt,minimum size=6mm},
%  link/.style={rectangle,draw=red!50,fill=red!20,thick,inner sep=0pt,minimum size=8mm}]
% \node (a1) at (0, 1) [server] {};
% \node (a2) at (0, -1) [server] {};
% \node (a3) at (-1.5, 0) [server] {};

% \node (b1) at (1, 1) [server] {};
% \node (b2) at (1, -1) [server] {};
% \node (b3) at (2.5, 0) [server] {};

% \node (shared) at (0.5, 0) [link] {Link};

% \draw [->] (a1) to [out=-60,in=60] (a2);    
% \draw [->] (a2) to [out=180,in=-60] (a3);
% \draw [->] (a3) to [out=60,in=180] (a1);

% \draw [->] (b1) to [out=240,in=120] (b2);
% \draw [->] (b2) to [out=0,in=240] (b3);
% \draw [->] (b3) to [out=120,in=0] (b1);

% \end{tikzpicture}
%     \caption{Two ring all-reduce jobs sharing a link}
%     \label{fig:mlsys-2rings}
% \end{figure}

\begin{figure}[!t]
\centering
\begin{subfigure}{0.23\textwidth}
    \centering
    \includegraphics[width=0.9\textwidth]{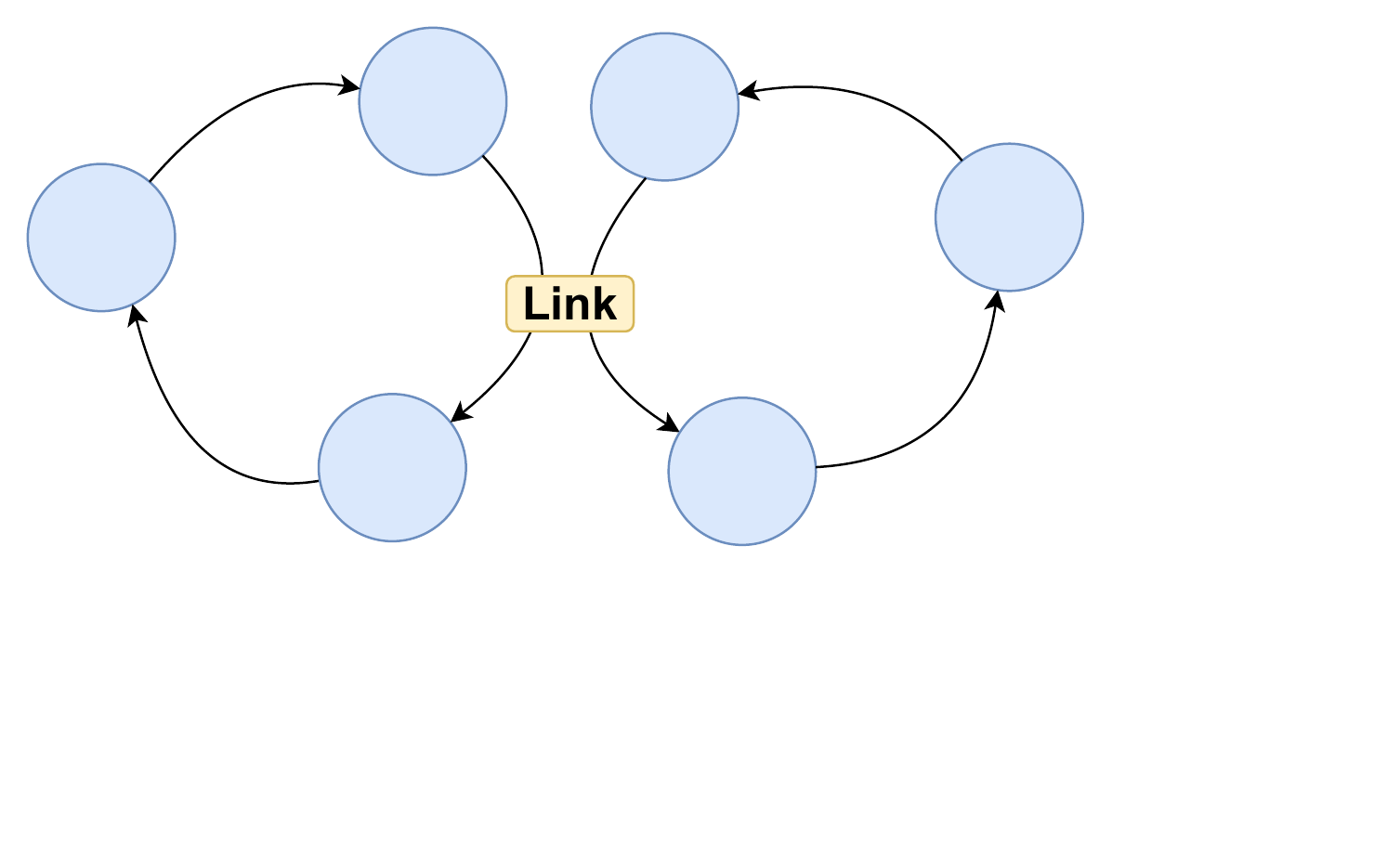}
    \vspace{-0.25in}
    \caption{}
    \vspace{0.1in}
    \label{fig:r4}
\end{subfigure}
\begin{subfigure}{0.23\textwidth}
    \centering
    \includegraphics[width=0.9\textwidth]{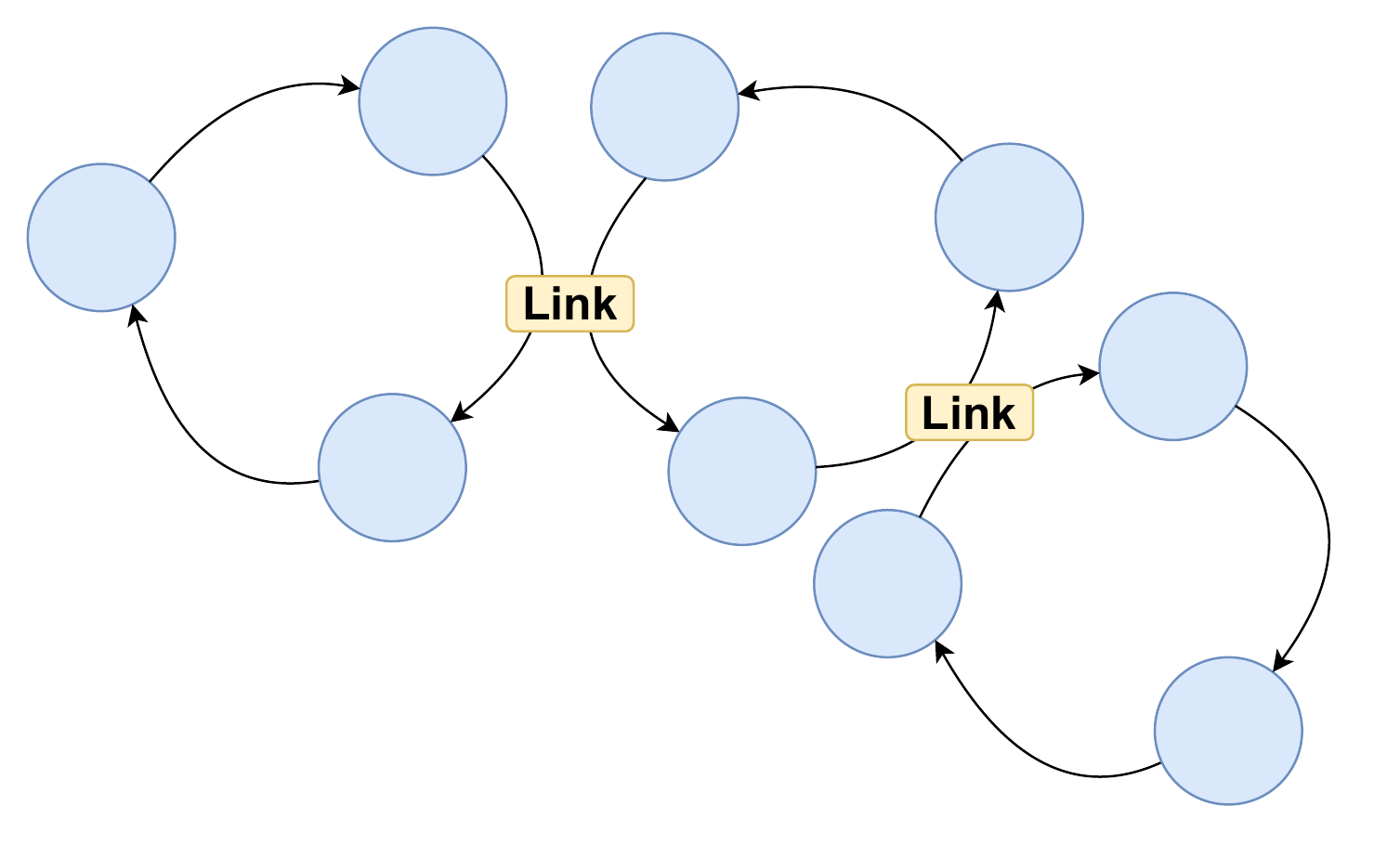}
    \vspace{-0.25in}
    \caption{}
    \vspace{0.1in}
    \label{fig:r1}
\end{subfigure}
\begin{subfigure}{0.18\textwidth}
    \centering
    \includegraphics[width=0.9\textwidth]{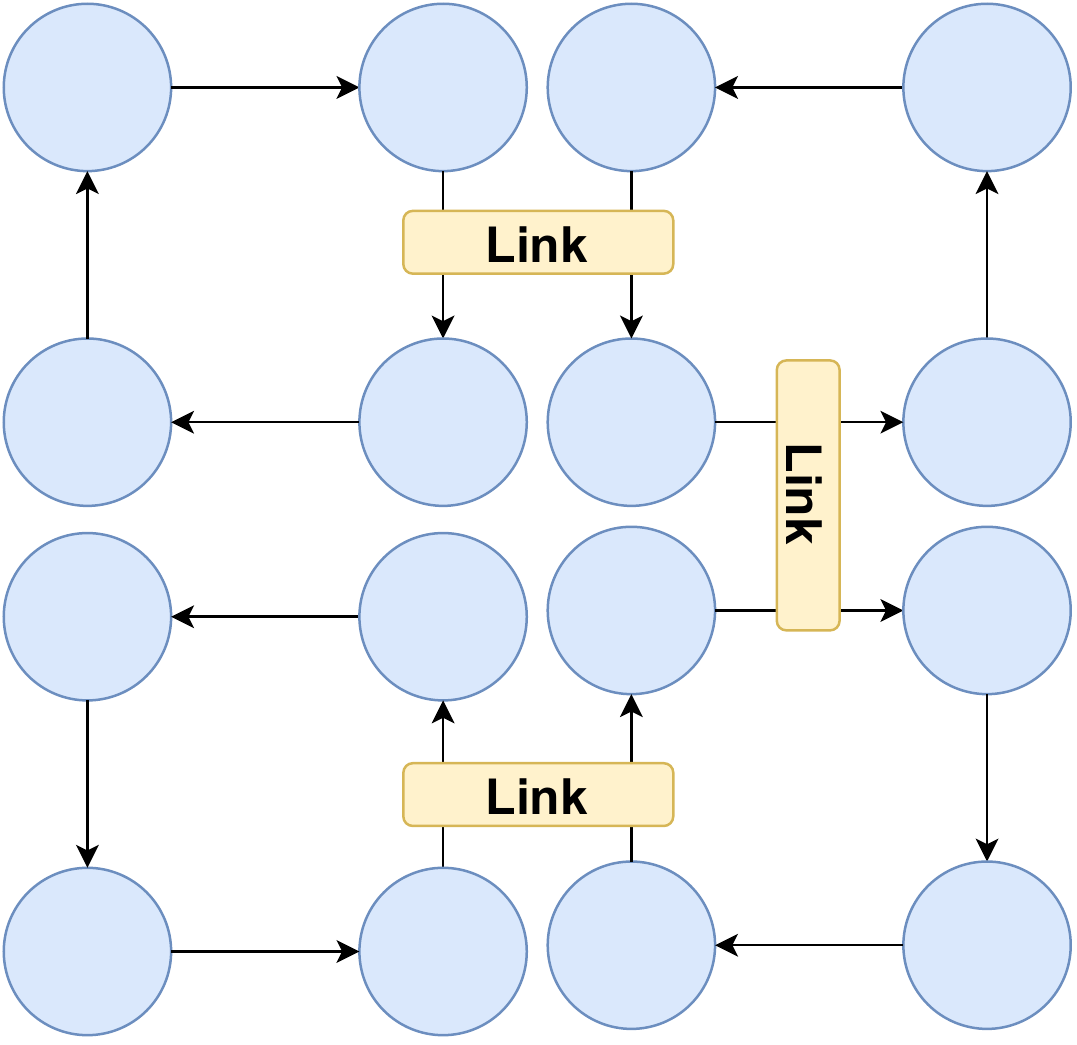}
    \vspace{-0.1in}
    \caption{}
    %\vspace{-0.1in}
    \label{fig:r2}
\end{subfigure}
\begin{subfigure}{0.23\textwidth}
    \centering
    \includegraphics[width=0.9\textwidth]{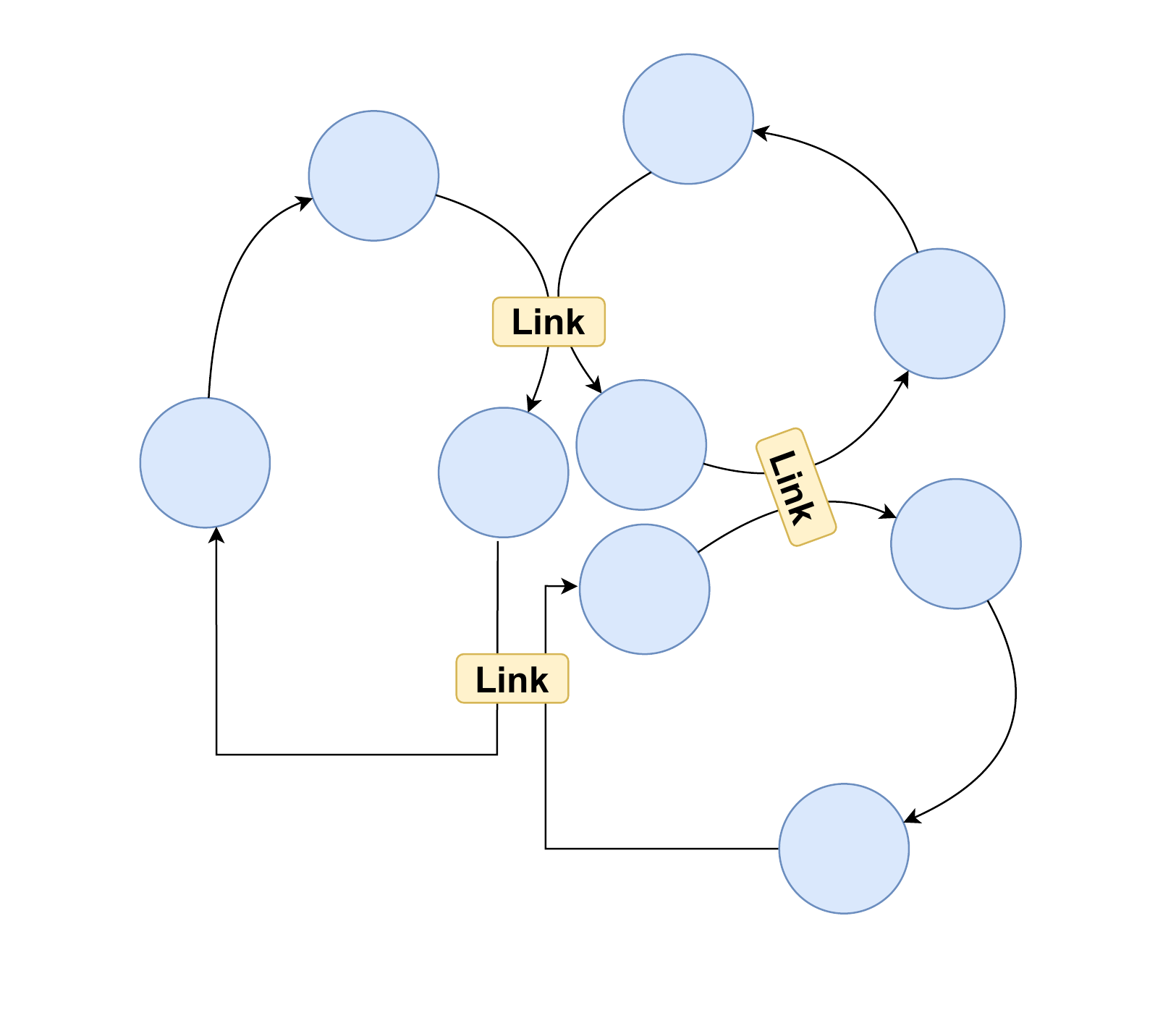}
    \vspace{-0.1in}
    \caption{}
    %\vspace{-0.1in}
    \label{fig:r3}
\end{subfigure}
\caption{Configurations of ring reduce jobs sharing links}
\vspace{-0.15in}
\label{fig:rr}
\end{figure}

%The above examples look at well-known heuristics. Perhaps the most compelling use case for our methodology is to understand less well-studied systems. 
Next, we consider the story of a recent paper~\cite{ml-net-sync} that examines complex emergent behavior in a cluster running multiple distributed neural network training jobs. Neural network training processes data in batches, computing the gradient of the error function for each batch to adjust the weights before processing the next batch. The computation-communication pattern for each batch is similar, making the process predictable and providing opportunities for better scheduling~\cite{byteps_2,muri,gandiva}. 

A training job consists of $n$ servers connected logically in a ring. During training, servers have phases involving intense computation followed by communication along the ring. A datacenter may run multiple jobs sharing the same physical network. As a result, some of the physical links may be shared by multiple training jobs as shown in Figure~\ref{fig:rr}. One can maximize network utilization by scheduling one job's computation while the other communicates so that each job gets the full available bandwidth when it is communicating. The authors of the reference paper~\cite{ml-net-sync} were surprised to find that a system that scheduled the jobs in this way did not perform much better than the vanilla one. Upon investigation, they uncovered that even in the vanilla system, jobs were spontaneously synchronizing to this desired schedule due to TCP unfairness. In this section, we verify this claim under more complex settings and find it to be true.

%Figure~\ref{fig:rr-intuition} gives intuition for why synchronization occurs. 

To understand why synchronization occurs, suppose job 1 has been transmitting on the shared link at full link capacity, and a new job 2 starts transmitting over the same link. Both jobs will detect the congestion and adjust their sending rates so that both transmit at half of the full capacity. However, it takes time for the transmission rates to reach this new equilibrium and many congestion control algorithms never reach it. In the meantime, job 1 gets more bandwidth than flow 2, so the job corresponding to flow 1 will finish its sum and broadcast steps earlier than it otherwise would have, which will make it start even sooner for the next batch. This will continue until the transmissions on the shared link are fully separated in time -- exactly the ideal schedule. Appendix~\ref{app:rr-vis} gives a visual representation of synchronization. 

This argument works when there are just two rings. But what if there are more? One shared link could be forcing the job to slow its schedule in one direction, while another shared link has an opposite effect. There could even be cycles.

%Each round of training processes one batch and consists of three phases: backpropagation, sum, and broadcast. The model is divided into $n$ pieces. Servers transmit data as soon as it is available. When summing, the first piece is available as soon as backpropagation is done processing \emph{that} part of the neural network. Then, the server must wait for data from the preceding server before forwarding it. The same holds for the broadcast step, except that the first piece is immediately available.

To verify synchronization on these more complex topologies, we use \framename to build a model of the training system. We provide a detailed description of our \framename-based model of Ring Allreduce Scheduling in Appendix~\ref{app:rr}, and here provide a summary of our findings. We ask the solver to find a case where overlap in communication increases (or remains constant), subject to a constraint that the communication time of each job fits inside the computation time of its neighbors. We find that the prior result does generalize to to the more complex topologies shown in Figure~\ref{fig:rr}:. It is important to note that the solver only proves this for a small and finite number of states. But because the initial state is unconstrained, this also proves that a larger sequence of events cannot exist where communication overlap increases; if such a sequence existed, it would have a short increasing sub-sequence which would be detected by the solver.

%, since this is the only case in which the results in our reference paper hold.  If the solver cannot find such a case, we have proved that synchronization will always occur.
%\vspace{-0.1in}

\begin{comment}
\begin{center}
\noindent\fbox{%
    \parbox{.9\linewidth}{
    \framename verifies a recently observation that TCP unfairness can cause ML training workloads to spontaneously converge to a state of high network utilization.
    }}
\end{center}
\end{comment}

\subsection{Replication of Prior Work}
In addition to the four novel case studies presented in this paper, we evaluate \framename's generality by using it to replicate the results of two prior works on performance verification: CCAC~\cite{ccac} and FPerf~\cite{fperf}. We provide the details of our model for these two examples in Appendices~\ref{sec:ccac} (congestion control) and~\ref{sec:fperf} (packet scheduling). We were able to confirm the results of both existing works with \framename in only one week.

\section{Limitations and Conclusion}

%\bm{Table copmaring case studies -- including at least No per-task variables/per-task variables}

%\bm{\textbf{Discussion: why not a compiler yet?} The sticking point is that we don't have a way to generate SMT-LIB versions of assign and similar functions. Second }

%While we believe \framename to be generally applicable to all scheduling algorithms, it is not a panacea for detecting problems in an algorithm. \framename still depends on users' ingenuity in mapping problems to our framework, then encoding their mapping in logical formulas. Further, it's still a burden on the user to formulate queries and understand the traces generated by the solver. 
%Our modeling approach can also only find performance issues for which a user formulates a metric and query; deeper issues which are not exposed by user--generated queries are not detected by our approach. Solver performance limitations mean that we can only perform bounded model checking, so \framename won't detect problems related to large numbers of tasks or events. It's still up to the user to further optimize the overall encoding of their system to enable the verification of larger problems (e.g., more tasks). However, we find that in many scenarios, it's possible to generalize our results by creating models that focus on critical system components and formulating reasonable queries. 

In this paper, we have demonstrated that formal methods can provide a deeper and more rigorous understanding of scheduling heuristics used in practice. Further, since we verify the \emph{specification} of these heuristics, and not the code, verification effort is minimal. The most time consuming part of our method is posing the right queries and interpreting counterexamples in context. For example, our model of the Linux CFS load balancer included several simplifying assumptions (e.g., ignoring asynchronous load balancing steps), yet the model was useful enough to detect practical bugs.  We invite the community to adopt performance verification as a part of their worklflow when developing scheduling heuristics.

This work does not raise any ethical issues.

\newpage
\bibliographystyle{plain}
\bibliography{cite}
% \newpage

\input{appendix}

\end{document}

%% file: 01_introduction.tex
\section{Introduction}

Modern software systems employ many heuristics to schedule resources such as CPU cores, virtual machines, and network paths. They undergo continuous development and fine-tuning to cope with the diversity of system characteristics and workloads. Evaluating their performance is a key challenge. Simulations are easy to do, but may not be realistic. Evaluation in deployment is realistic but can require extensive development effort. Further, instrumenting a real testbed to get sufficient visibility while emulating realistic workloads can be non-trivial and add significant overhead~\cite{CrystalNet}. 

Recently, there has been interest in using methods from formal verification to evaluate the performance properties of heuristics~\cite{ccac,fperf,ccmatic,te-gap,cc-fuzz,metaopt}. Approaches like these are examples of \textit{performance verification}: the use of solvers, theorem provers, or other formal tools to provide theoretical guarantees about the performance of an algorithm on a particular system. %Where classical formal methods research focuses on questions of correctness, performance verification makes queries about efficiency. 
Performance verification overcomes a key limitation of simulators. In particular, simulations can only check specific workloads manually selected by developers, increasing developer effort and potentially leaving out significant behaviors. Formal methods can automatically reason about combinatorially many workloads and system designs, providing thorough and conclusive answers about which conditions fall under the assumptions made by a model. Prior work on formal verification of performance properties focuses on specific individual domains: congestion control~\cite{ccac,arun2022starvation}, packet scheduling~\cite{fperf}, and traffic engineering~\cite{metaopt,te-gap}. In these domains, verification approaches have discovered previously unknown performance issues and proven performance guarantees. However, these approaches require considerable human ingenuity and effort which must be replicated by hand for each domain. Developing new techniques for each of hundreds of diverse heuristics would be infeasible, particularly for the long tail of less well-studied examples. 

%\aditya{so far the intro seems to use heuristic and system in confusing ways. Are prior works about systems or heuristics? What does it mean for a heuristic to impact a system?}\bm{fixed}
This paper seeks to help the typical system designer with minimal formal methods background  to obtain \emph{useful} and \emph{actionable} insights about how their heuristic of interest will impact their system. To that end, we conduct performance verification on six different heuristics used in six different systems. Through our experience, we identified two fundamental challenges and propose a framework, \framename, that helps overcome them. First, since verification performs worst-case analysis, it can be overly pessimistic. We offer guidance on how to discover a \emph{minimal} set of assumptions that the designer needs to make to obtain useful insights. Second, real systems are extremely complicated, which makes it impractical for humans to fully model them and for computers to reason about the resulting models. We recommend that the designer break down the heuristic's execution into a small and linear sequence of steps, over-approximating when necessary to preserve rigor while reducing model complexity.

Our six case studies illustrate \framename. The first four are new and four distinct students worked on them sequentially, evolving \framename with each use-case. The last two replicate prior work using our framework~\cite{fperf,ccac}. The four case studies investigate 1) work stealing scheduling, 2) the Linux CFS load balancer~\cite{fair.c, v5.5-lore}, 3) single processor shortest remaining time first scheduling (SRPTF), and 4) a recent network scheduler for neural network training clusters~\cite{ml-net-sync}. The first and third case studies took time as we were still developing the methodology. In the second, an undergraduate student, already knowledgeable about the Linux load balancer, was able to discover bugs in three weeks. In the fourth, a PhD student already experienced in formal methods and systems, extended an existing result on convergence to new workload classes in just three days. Finally, we were able to replicate some of the findings from FPerf~\cite{fperf}\footnote{To aid human interpretation, FPerf produces an entire class of workloads that breaks a given heuristic. Our methods produce only a single workload and are hence simpler. For our case studies, this is sufficiently interpretable.} and CCAC~\cite{ccac}\footnote{CCAC developed a custom network model, which we capture using Virelay.} in one week. Thus \framename enables designers to \emph{quickly} obtain actionable insights about heuristics used in real-world systems.

While our primary contribution is the framework, our four case studies also find new and actionable insights about real-world heuristics that are independently interesting. We will release the code of \framename along with a comprehensive user guide after the publication of this paper. %This work does not raise any ethical issues.

%In summary, we make the following contributions:

%\noindent \textbf{1.} We study the design principles belaying performance verification and distill them into an easy-to-use framework that applies to a wide range of resource allocation heuristics.
%and leverage that understanding to develop a set of easy-to-follow guidelines that are applicable to a wide range of resource allocation heuristics.

%\noindent \textbf{2.} We demonstrates the generality and ease of use of \framename by applying it to four use cases which have not previously been subjects of performance verification. We also replicate results from prior performance verification work.

%\noindent \textbf{3.} We provide actionable insights about the algorithms studied in each use case, including identifying an algorithmic bug in the Linux CFS load balancer.

%\noindent 1. Presents a general and easy-to-use framework for applying formal methods to studying performance properties of resource allocation algorithms.

%\noindent 2. Demonstrates the generality and ease of use of the framework by applying it to four use cases that haven't been studied before with performance verification. We also replicate results from earlier work on performance verification.

%\noindent 3. Provides actionable insights about the algorithms studied in each use case, including identifying an algorithmic bug in the Linux CFS load balancer. 

%% file: 02_motivation.tex
\section{Motivation and Design Rationale}

\subsection{Scenario}

The CFS load balancer is an important component of Linux. It has seen continuous development for over two decades. % without reaching a consensus. This is understandable since it operates everything from microwave ovens to the largest super-computers. 
Ideally, the load balancer should be work conserving while avoiding significant load imbalance. Before 2019, it had several well-documented problems~\cite{wastedcores, v5.5-lore}, leading to a complete rework of its code in 2019~\cite{v5.5-lore}. Unsurprisingly, the rework release note stated that ``there is going to be some workload that is hurting from these changes''~\cite{release-note}. Indeed, shortly after the rework, it was discovered that imbalance can persist due to inaccurate accounting of the load on an overloaded CPU~\cite{rework-regression}. Several major adjustments have been made since the rework, based on a process of trial and error. In this process, performance regression is detected by a kernel expert for a workload they regularly deploy, leading to some discussion between maintainers and an eventual fix. Instances of this process are numerous~\cite{hackbench, linux-bench, rework-regression-2, fairness-report}. This process of trial and error is inefficient.

%The question arises: ``does a given update suffer from performance regression?'' Current approaches that rely on benchmarking are clearly insufficient. 

%\aditya{I found this paragraph redundant wrt intro and 2.2}
In this paper, we argue that performance verification can more efficiently and systematically detect such performance bugs. %(e.g., performance regression due to an update in the logic of a heuristic). 
%It does so by using formal methods to identify scenarios of poor performance anywhere in the whole state space of a system, rather than relying on the ad-hoc discovery of bug-exposing workloads. Performance verification addresses the shortcomings of the traditional trio of performance analysis tools: benchmarking, simulation, and theoretical modeling. Benchmarking a full implementation offers realistic measurements, but provides coverage for only the test cases included in the benchmark. Further, it is not always feasible to implement a full system to test a new resource allocation heuristic. While simulations might be lower effort, they still require accurate workload and system characterization. Theoretical analysis using queueing theory or control theory is not tractable without oversimplifying assumptions, making it of limited utility, especially for complex systems like the Linux load balancer. Performance verification, on the other hand, covers all possible input workloads while balancing modeling effort and tractability. It does so by leveraging recent advancements in automated solvers for satisfiability modulo theories (SMT) constraints~\cite{xxx, xxx, xxx}.%, capturing models of real systems while producing interesting results~\cite{xxx,xxx,xxx}. 
%In the rest of this paper, we demonstrate the power of performance verification to detect real performance bugs over highly complex systems. For instance, , 
We use our approach to reproduce existing performance bugs and identify new performance problems in the latest Linux CFS load balancer (i.e., v6.8-rc1). For example, we find that an overloaded CPU can have some of its tasks migrated to another CPU, only for that CPU to steal them back shortly after. The result is that the overloaded CPU remains the only overloaded CPU in the system, with its tasks receiving less than 60\% the execution time of any other task in the system (\S\ref{sec:linux}).

%tasks can get stuck on an overloaded CPU because the load balancer doesn't track the history of execution of a task that has been mi

%\ahmed{modify based on Jehad's results}

Despite its advantages, performance verification for systems remains a nascent area, and the tools thus far are bespoke, tailored for the particular system-heuristic pairs they analyze. Developing a custom model of a heuristic like the Linux CFS load balancer, consisting of 10,000 lines of code, is a tremendous undertaking, requiring expertise spanning kernel development, modeling, and formal verification methods. Producing bespoke models from scratch for every complex heuristic would therefore be impractical. In the following sections, we investigate the principles underlying performance verification approaches and use that understanding to develop a general framework to guide developers in performance verification of new systems.

\subsection{Assumption-Constrained Worst-Case Analysis}
%Performance verification for systems is a nascent area and the tools thus far are bespoke, tailored for the particular systems they analyze. To make the approach more general, we have to identify the 

The driving insight behind performance verification is that worst-case analysis can account for real-world behaviors that probabilistic analysis can miss. 
Consider why developing the Linux CFS load balancer has been such a confusing enterprise. Its analysis has been largely probabilistic; developers run benchmarks and measure the performance. However, it is very difficult to generalize these results; every change to the heuristic helps some workloads while hurting others. The difficulty arises from the fundamental problem that probabilistic analyses make too many assumptions about the workload and environment. Every benchmark embodies very specific distributions for the number of threads, their relationships to each other, and their blocking characteristics. Individual benchmarks can therefore miss real-world circumstances where the heuristic fails.

%has very specific distributions of where and how many threads are created, when and for how long they block, and how the threads communicate with each other. 
%Our community lacks the language to precisely state the assumptions the load balancer makes or the performance properties it provides.  

\noindent\textbf{Minimal assumptions} Worst-case analysis addresses the limitations of benchmarking by considering \emph{all} the possible circumstances a heuristic may face. In essence, performance verification describes a system and then lets an ``adversary'', i.e., a solver, choose among all possible inputs in a bid to produce undesired behavior. While simulations are made more realistic by adding detail to the models, performance verification is improved by \emph{removing} details. This is because, when we remove details, we reduce the number of constraints in the model, which allows the solver to adversarially choose among more possible behaviors.  The tradeoff is that this approach is \emph{overapproximative}, a feature common to many types of formal verification. In an overapproximation, any behavior of the real system is possible, but behaviors possible in the model may not occur in practice. Thus performance verification with minimal assumptions may be too \emph{pessimistic}.

To combat this drawback, we usually need \emph{assumption-constrained} worst-case analysis. Assumptions often arise when the current state of the system or its future inputs are unknown or too expensive to determine. A benchmark of the Linux load balancer makes assumptions about factors like 1) the distribution for the CPU and IO requirements of each task which determines how much CPU load a task will create, 2) the cache access patterns of tasks which determines their locality, 3) the behavior of the CFS scheduler which impacts fairness between tasks executing on the same CPU.  During performance verification, we achieve worst-case analysis subject to constraints by allowing the ``adversary'' to decide the initial values for the parameters of all tasks, e.g., blocking and running times, allowing it to explore all possible task behaviors. If the property we want to verify is true in spite of an adversarial assignment of inputs, it will be true in any real system. If the property is false, users can inspect how the adversary broke the property, revealing what restrictions are necessary to make the property true. This exercise {\em reveals which assumptions about a system's environment are implicitly embedded in a heuristic}. For example, assuming that the CFS scheduler can be unfair to tasks scheduled on the same CPU necessarily leads to unfairness between tasks that the load balancer cannot fix. Thus, we add to our model the assumption that the CFS scheduler is fair.

Because the constraining assumptions used in performance verification are {\em minimal}, the size of the explored input space remains far larger than that exercised by benchmarking or simulation. However, even with an iterative process of adding assumptions, developers face a challenge in organizing the assumptions they make, and in determining whether these assumptions are minimal. In the next subsection, we develop a guide to help developers respond to these challenges.

\noindent \textbf{What questions can it answer?} Since performance verification does not consider concrete distributions, it cannot answer quantitative questions like ``what is the $99^{\mathrm{th}}$ percentile latency''? Instead, performance verification is to benchmarks and simulations what asymptotic analysis of an algorithm is to measuring its running time. It answers qualitative questions about the performance of a heuristic such as ``is the process scheduler eventually work conserving?'' (see~\S\ref{sec:linux}), ``does the distributed flow synchronization mechanism eventually converge to a good state?'' (see~\S\ref{sec:rr}), or ``does this congestion control algorithm eventually achieve non-zero utilization?'' (see~\S\ref{sec:ccac}). In addition to these questions, performance verification can answer quantitative questions about worst-case performance, such as comparing a heuristic's decisions to an offline optimal solution (see~\S\ref{sec:srpt} and \cite{metaopt}).

\subsection{Human-computer Collaboration}

Humans can find it hard to search through the exponentially many possible execution paths of a heuristic. This is particularly true for control algorithms that operate over multiple timesteps, since the temporal dynamics may be counter-intuitive for humans. Automated reasoning with SMT solvers~\cite{z3,cvc5} can find insights that humans miss. However it is unreasonable to expect a computer to automatically analyze a complex system. Humans are much better at abstracting to the appropriate level. \framename provides recommendations on how to best combine the strengths of human and computer reasoning.
% Outline:
% - Small number of discrete events
% - Encoding tricks and over approximation
% - How to refine questions?
As with any evaluation of a heuristic including simulations and real-world testing, we recommend that the model designer begin by identifying 1) the heuristic algorithm, 2) the system it is operating under, and 3) the performance properties they wish to monitor. \framename is built around the following ideas.

%\noindent\textbf{Model components.} 

\begin{figure}
\centering

\resizebox{\columnwidth}{!}{
    \begin{tikzpicture}
    \tikzset{
    partial ellipse/.style args={#1:#2:#3}{
        insert path={+ (#1:#3) arc (#1:#2:#3)}
    }
}

        \node[draw,minimum height=0.6cm, minimum width=1.2cm] (A) at (0,0) {\texttt{state}$_0$};
        \node[draw,minimum height=0.6cm, minimum width=1.2cm] (B) at (2,0) {\texttt{state}$_0'$};
        \node[draw,minimum height=0.6cm, minimum width=1.2cm] (C) at (4,0) {\texttt{state}$_1$};
        \node[draw,minimum height=0.6cm, minimum width=1.2cm] (D) at (6,0) {\texttt{state}$_1'$};
        \node[draw,minimum height=0.6cm, minimum width=1.2cm] (E) at (8,0) {\texttt{state}$_2$};
        \draw [->] (A) -- (B) node[pos=0.5,above,yshift=8,xshift=10,rotate=45] {\tiny{\texttt{Algorithm}}};
        \draw [->] (B) -- (C) node[pos=0.5,above,yshift=4,xshift=10,rotate=45] {\tiny{\texttt{System}}};;
        \draw [->] (C) -- (D) node[pos=0.5,above,yshift=8,xshift=10,rotate=45] {\tiny{\texttt{Algorithm}}};;
        \draw [->] (D) -- (E) node[pos=0.5,above,yshift=4,xshift=10,rotate=45] {\tiny{\texttt{System}}};;

        \node[draw,minimum height=0.4cm, minimum width=1.9cm] (F) at (4,-0.8) {\small{\texttt{Algorithm}}};
        \node[draw,minimum height=0.4cm, minimum width=1.9cm] (G) at (4,-2) {\small{\texttt{System}}};
        \node[draw, ellipse] (H) at (5.5,-1.4) {\tiny{\texttt{State}}};
        \node[draw, ellipse] (I) at (2.5,-1.4) {\tiny{\texttt{State}}};

        \draw [->] (F.west) arc (90:160:0.5);
        \draw [->] (I.south) arc (180:270:0.4);
        \draw [->] (G.east) arc (270:340:0.5);
        \draw [->] (H.north) arc (0:90:0.4);

\end{tikzpicture}
}
\caption{Overview of alternating calls between \texttt{System} and \texttt{Algorithm}.}
\label{fig:model_diagram}
\end{figure}

\noindent\textbf{Small and linear sequence of events.}  We advocate following the modeling approach used in discrete-event simulation. The designer expresses interaction between the system and the heuristic with a \emph{small} and \emph{linear} sequence of alternating calls between the two (see Figure~\ref{fig:model_diagram}). The heuristic observes the system state and makes a control decision. The system then updates its state in response to the decision. We call a heuristic invocation an ``event''; state changes occur because of heuristic action (at the event) and because of the operation of the system (between events). Under this model, both the system and the heuristic are modeled as simple state transition functions. We advocate for this approach for two reasons.

%Identifying the unique interactions between the heuristic and the system for each modeled heuristic is a laborious task. 
%We refer to each call as an ``event''
%With each state transition considered an ``event,'' this model resembles models used in discrete-event simulations, a common approach in simulating networked systems, making it an approachable way for designing models. 
%In addition to providing a simple template for modeling, there are two more reasons we advocate for this approach. 

First, having a {\em small} sequence is necessary to make fully-automated verification feasible. Further, we show that a  small sequence is  sufficient to capture interesting behaviors in many heuristics. There are two ways to ensure that a small sequence of events is representative of larger realistic sequences. The rigorous method is to use the SMT solver to prove lemmas over short event sequences, and use them in hand-written proofs that apply to arbitrarily long sequences of events using the principle of mathematical induction. The hand-written parts of the proofs are easy to write, since the SMT solver is doing the heavy lifting. Two of our case studies, in sections~\ref{sec:rr} and~\ref{sec:linux} (and one prior work~\cite{ccac}), use this method. There is another non-rigorous approach that is used by our other two case studies. Here, we simply perform bounded model checking and assume that the results apply to larger sequences. Our confidence in such assumptions comes from the highly regular structure of the output (see~\S\ref{sec:ws} and~\S\ref{sec:srpt}).

%Second, since humans tend to think imperatively, it is easier to write constraints when we have a linear sequence. 

Second, the linear structure makes the constraints shorter and simpler to write. In particular, the template of a discrete-event simulator allows for encoding the model as a sequence of identical invocations of the heuristic and system, differing only in their input (i.e., system state). At each event, we replicate all the state so that the state at time $t$ only depends on the variables at time $t-1$. 

%\aditya{Drop the rest. It doesn't add much and is confusing} Our earlier attempts avoided this duplication in hope of making the SMT solver's job easier. However this made the constraints convoluted and the overall formulation buggy. Our experiments show that the SMT solver is just as fast at solving constraints written using our recommended (see~Figure~\ref{fig:wsa}).

%In earlier versions, we were tempted to write constraints based on explicit modeling of discrete time steps, thinking this approach would make the solver faster \bm {I'm not sure we actually thought this, and it doesn't make intuitive sense. This paragraph has a different conclusion than I thought when collecting data}. In reality, this approach made the encoding more convoluted while actually hindering performance; for example, the \framename--based model of SRPT scheduling can solve a query on the number of tasks which can be finished by a deadline within $52$ seconds, while our naive model does not produce a result within one hour. \bm{maybe we can still add from the other case studies?}

\noindent\textbf{Iterative refinement of model assumptions.} Overapproximation is a necessary component in performance verification. We advocate that models primarily overappoximate the behavior of the system, not the heuristic. In particular, the model designer should attempt to capture the heuristic with high fidelity, as its the focus of the modeling effort. On the other hand, the system can be too complicated for humans to model with high fidelity. Thus, we conversely advocate that the model of the system be as loose as possible, adding only the bare minimum constraints on system behavior, relying on overapproximation as discussed earlier. For instance, in work stealing we start by allowing execution time and context switching costs to be arbitrary real numbers, with the only constraint being that they be positive. We can retain confidence in the minimality of assumptions by developing them through an iterative interaction between a developer adding assumptions and the adversarial solver exposing the need for new ones.

%\aditya{drop the rest. completely redundant.} 
%Because worst case analysis is pessimistic, it will show execution traces where the heuristic's decisions are arbitrarily bad. %This is because we are looking at heuristics that are fundamentally information-constrained, and \emph{cannot} always make perfect decisions. Performance verification allows designers to discover minimal assumptions under which we prove a performance property.

To enable designers to discover minimal assumptions under which we can prove a performance property, \framename gives the designer solver outputs representing a small example where the heuristic fails by an extreme amount. In work stealing, for instance, such an example might be a 3-task job where the performance of work stealing is $100\times$ worse than optimal. These simple examples make the necessary conditions for poor performance very clear. Moreover, they can be generated iteratively: a designer may use an example to find a missing assumption, add that assumption, and then generate another example that illuminates further necessary assumptions. At the end of this process, the user is left with a model that makes the smallest possible set of assumptions while still providing useful results. We have found this iteration strategy to be a key benefit of \framename in several case studies. In work stealing, for instance, it allowed us to discover that the ratio of context-switching costs between tasks is a key determinant of performance.%, while in the case of SRPT it led to the discovery of several bugs in our initial implementation of the \sys function.

%While assumption discovery is challenging to automate, a human looking at a \emph{small} and \emph{extreme} example of failure will usually find the assumptions quite easily.

\subsection{Limitations}

%\aditya{just call this Limitations and drop the "compiler" stuff. It is distracting.}
%\ahmed{A discussion of generality and ease of use might be warranted here.}

%\noindent \textbf{Limitations.}
Naturally, \framename is not a panacea. \framename helps model a single deterministic %and relatively abstract 
heuristic in a system that can be abstracted as a single transition function. In other words, \framename does not support modeling the interactions between different heuristics that jointly control a system as it would require identifying all possible interleavings of the invocations of the contollers and the system. In scenarios where multiple heuristics interact, we combine them in a single controller, if possible, or treat all but one of them as part of the system. Further, \framename doesn't support modeling randomized algorithms nor is it suitable for verifying the implementation of a heuristic. Like existing tools, \framename requires its users to encode a separate representation of the heuristic~\cite{ccac,fperf,metaopt}. Like other model-checking-based techniques, \framename only supports relatively small instances of the modeled problem, which we found to be sufficient to prove interesting properties.  Finally, \framename is built on a one-size-fits-all template which is bound to increase the model size compared to bespoke models. Yet, in our experiments, verification time was almost identical for the \framename model of work stealing and the bespoke model we developed for it. 

%However, we found it to be sufficiently efficient to verifying for the model sizes we picked.

%However, in our experience modeling a few timesteps is usually enough to verify the performance predicate of interest or show a useful counterexample that identifies a performance bug. \ahmed{mention that creating a template is bound to be less efficient that a bespoke model, we probably need to highlight the gap with results}

%\noindent \textbf{Why not a compiler, yet?} \framename still requires a considerable amount of manual labor from its users to model their algorithms and then encode parts of their model in logical formulas. \framename's objective is to simplify that process, making it accessible to a broader range of users. Ideally, we should be able to automate the process of converting an imperative description of the algorithm to SMT formulas. Such a compiler should also be able to optimize the SMT encoding, making it faster to verify. However, we believe that at this stage in the development of performance verification tools, it's still favorable to err on the side of generality, where a user is allowed to encode their algorithm freely. An efficient compiler might either limit the flexibility of the framework by providing restrictive abstractions or provide abstractions that are too generic. We believe that \framename already provides a good balance between generality, ease of use, and efficiency. As this research area evolves to tackle more complex heuristics and systems, we anticipate the emergence of effective compiler abstractions.

\subsection{Related Work}

A long history of theoretical analysis of scheduling algorithms can be traced back to Graham's work on multiprocessor scheduling for jobs with precedence~\cite{graham}. Since then, a diverse class of related problems have been studied, including different machine environments (single, multiple-uniform, multiple-unrelated), job characteristics (processing times, arrival times, preemption), and objective functions (mean/max completion time, throughput, sojourn time)~\cite{brucker-sched, brucker-sched2, pinedo-sched}. We focus on approaches based on formal methods and worst-case analysis. CCAC \cite{ccac} and FPerf \cite{fperf} build formal models for congestion control and packet scheduling algorithms. Our approach has been directly inspired from these principles of building abstract models with specifications to verify them. FPerf provides a tool that allows for synthesizing generalized traces that cause poor performance. FPerf's synthesis tool solves an important but complementary problem to \framename. Both FPerf and CCAC require expertise and significant human ingenuity, and cannot be easily applied in other settings. In contrast, we attempt to provide a general methodology that can be adapted to formally model a much larger class of schedulers. MetaOpt~\cite{metaopt}, like \framename, allows identifying the gap between a heuristic and an optimal algorithm. MetaOpt focuses on cases where the optimal algorithm and the heuristic can be defined precisely as optimization problems, providing an efficient solver for cases where the search space is convex. \framename can answer more elaborate questions about the performance of a heuristic. Further, \framename allows for analyzing performance over multiple events, while MetaOpt is concerned with a single invocation of the heuristic. However, MetaOpt can analyze larger problem instances than \framename.

\framename differs from existing work on Worst-Case Execution Time (WCET), applying formal methods to identify WCET scenarios for real-time systems. While WCET is concerned with identifying WCET under a completely specified model of a system, \framename is concerned with assumption-constrained worst-case analysis. Thus, \framename helps uncover a wider range of performance bugs by iteratively changing the assumptions added to the model. Further, \framename provides a general framework for analyzing heuristics. On the other hand, WCET analyses are all individually tailored to their specific systems~\cite{pinedo-sched, wcet-mp, wcet-mp2, ws-smt}.

%% file: 03_virelay.tex
\begin{figure}[!t]
    \centering
    \includegraphics[width=0.95\linewidth]{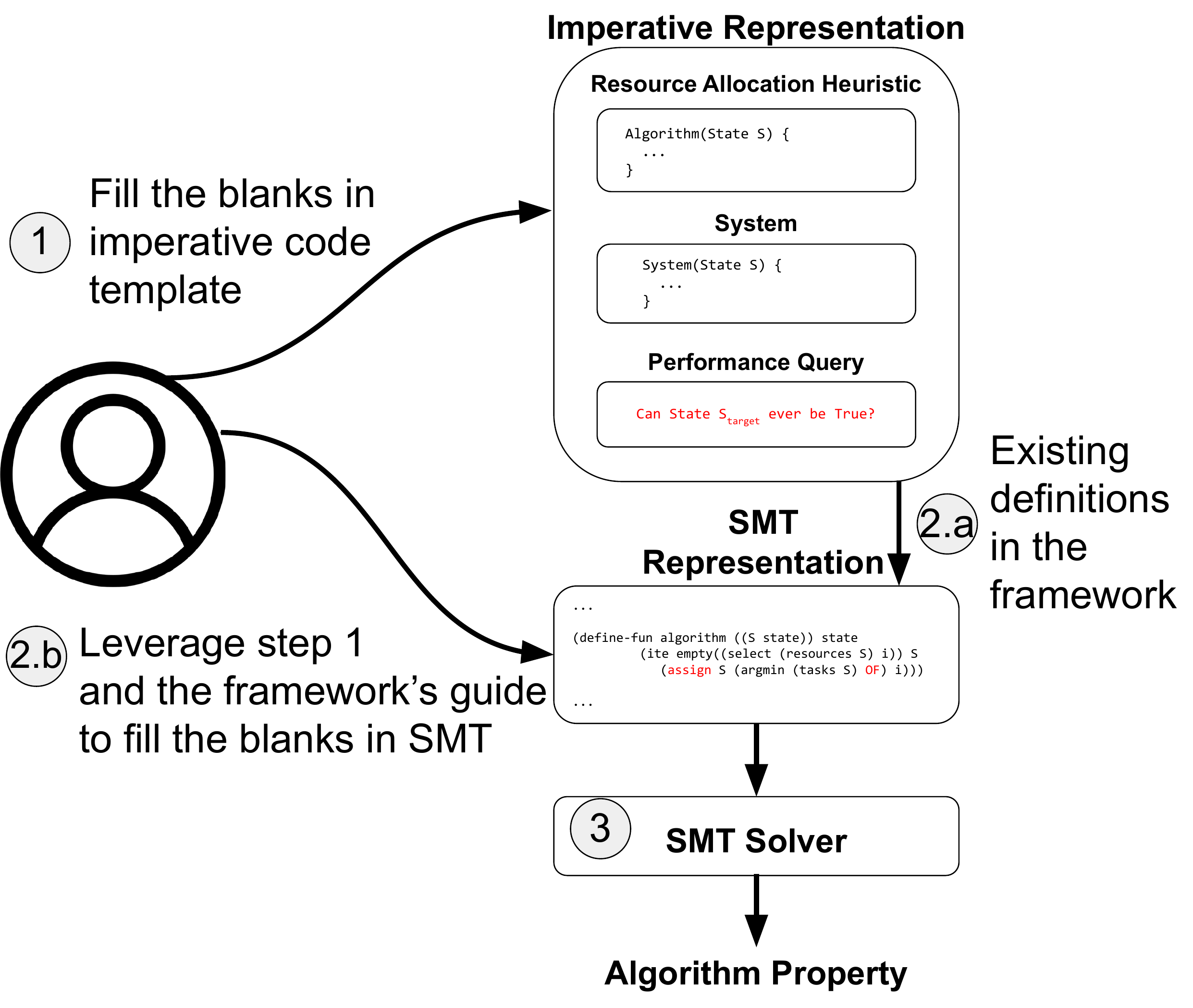}
    \vspace{-0.15in}
    \caption{Performance verification workflow under \framename. Although formal methods experts can jump right into step 2.b with little help from the guide or even existing definitions, we find that this framework helps formal methods novices develop useful models rapidly.}
    \label{fig:framework_flow}
    \vspace{-0.15in}
\end{figure}

\section{\framename}\label{s:methodology}

\label{sec:guidelines}

\newcommand{\true}{\textproc{true}\xspace}
\newcommand{\false}{\textproc{false}\xspace}
\newcommand{\bxt}[1]{\boxed{\texttt{#1}}}

In this section, we describe \framename, which allows users to bridge the gap between imperative system definitions and declarative verification constraints. Figure~\ref{fig:framework_flow} shows the workflow of a heuristic designer using \framename. First, they encode the behavior of their algorithm and system in imperative code within the skeleton provided by \framename. Designer can also encode queries about the performance of their algorithm which typically take the form of declarative statements. Then, they map their imperative code to SMT constraints using the guide we provide next in the paper. Components of \framename are already encoded as SMT constraints. Once the encoding is complete, the solver can find answers to performance queries made by the designer.

Any resource allocation problem involves some workload which we can represent as a set of tasks $T$, and a set of resources $R$ to be used by tasks. Tasks come with some properties representing the amount of work to be done and dependencies on other tasks. A resource is equipped with a queue that contains the tasks that may use that resource. 

\begin{figure}[!t]
    \centering
    \includegraphics[width=0.98\linewidth]{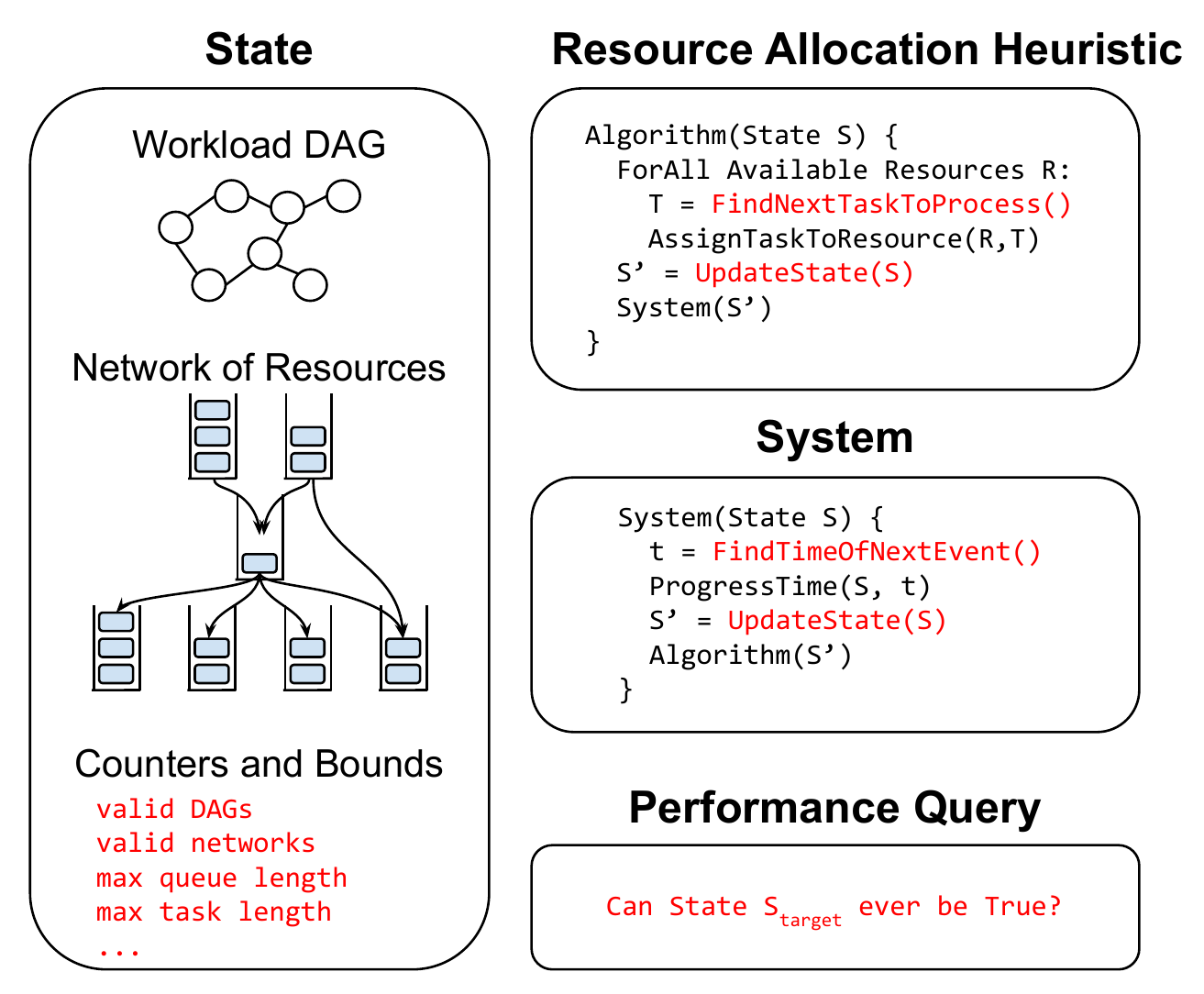}
    \vspace{-0.15in}
    \caption{An overview of \framename. A user only needs to fill in the red parts of the framework. More details about the exact structure of the framework are given in listings 1--4.}
    \label{fig:framework}
    \vspace{-0.15in}
\end{figure}

The start and end of work on a particular task by a particular resource are events associated with points in time. In the most general sense, any allocation heuristic is an ordering of these events. To make the verification problem tractable, we introduce an abstraction, \texttt{state} which holds all information about tasks, resources, and global counters. Every state is associated with a particular time, allowing us to represent the ordering of events as a sequence of states and state transitions.

Unfortunately, state transitions between events are complex and generally specified with imperative code. To make it easy for users to convert such imperative descriptions to constraints, we define \emph{two} transition functions between system states:  \texttt{System}, representing state transitions caused by the system-wide behaviors, and \texttt{Algorithm}, representing state transitions caused by the invocation of the allocation algorithm. Thus, our model is composed of an alternating sequence of calls to \texttt{Algorithm} and \texttt{System}, as shown in Figure~\ref{fig:model_diagram}.  Furthermore, we provide a general skeleton for each transition function. % in both imperative and declarative code. 
Thus, by filling in problem-specific ``blanks'' in each outline, developers can easily obtain declarative functions in the language of SMT representing the state transitions. Given these functions, a solver can reason about the system under study with a simple sequence of state transition constraints $(=\;S'\;{\tt System}({\tt Algorithm}(S)))$ for each step moving from state $S$ to state $S'$. Figure~\ref{fig:framework} shows in more detail the components of a Virelay model: a three-part state abstraction, two functions \texttt{Algorithm} and \texttt{System}, and a performance query.

\subsection{State}
The state of any resource allocation system includes a description of tasks, resources, and possibly additional system-wide state. We represent system state as a tuple $\texttt{state} = (T, Q, W, C)$, where $T$ is a set of tasks, $Q$ is a set of queues, one for each resource, $W$ is a set of variables, and $C$ is a set of constraints which represent invariants on state, as described below.

\begin{tightitemize}
\item Task-wise variables: A task $t_i = \left(v_0, v_1, \dots\right)$ is a tuple of variables that represent the state of a task. For instance, a task may include variables to represent how much work has been done, or the time at which work started.
\item Resource-wise variables (queues): Each resource is equipped with a queue $Q$, which includes an ordered set of tasks and possibly some additional variables to represent, for instance, the resource's throughput. The queue contains the tasks which can be assigned to this resource.
\item System-wide variables: Some scheduling systems may require system-wide variables to represent the global state, for instance, the system counters with the number of tasks that have been finished.
\item Invariants: A set of constraints over features of state encoding bounds and ``makes-sense'' requirements, for example, maximum task length or limits on valid DAGs.
\end{tightitemize}
In some cases, parts of the system state can be inferred from others; for instance, the number of completed tasks could be represented as a function of the set of tasks. The question of which parts of the state should be explicitly assigned SMT variables and which should be computed from the rest is an implementation detail that may affect solver efficiency, but not correctness.

\begin{figure}[!t]

\begin{lstlisting}[mathescape=true, caption={General \texttt{System} transition function in imperative form. This function represents state changes from the passage of time and the completion of work.}, language=Python,belowskip=-8pt, label={lst:imperativeSystem}]
var S #state
def System():
  if $\bxt{doneCondition(S)}$:
    exit S
  else:
    td = $\bxt{nextEvent(S)}$
    $\bxt{update(S, S.time-td)}$
    wait(S.time-td)
\end{lstlisting}
    %$\bxt{updateResources(S, S.time-td)}$
    %$\bxt{updateCounters(S, S.time-td)}$
\begin{lstlisting}[mathescape=true, caption={Declarative version of \texttt{System} in the SMT-LIB language.}, language=Python,belowskip=-8pt, label={lst:smtSystem}]
(define-fun system ((S state) (t Real)) state 
    (ite $\bxt{doneCondition(S)}$ markDone(S) 
        (let (td (- (time S) $\bxt{nextEvent(S)}$)) 
            wait(S, $\bxt{update(S, td)}$, td))))
\end{lstlisting}

\begin{lstlisting}[mathescape=true, caption={General \texttt{Algorithm} transition function in imperative form. This function represents state changes from the application of a scheduling heuristic.}, language=Python,belowskip=-8pt, label={lst:imperativeAlgorithm}]
var S #state
def Algorithm():
  for j in S.R:
    if S.R[j].q empty:
        ta = $\underset{i \in {\tt S.T}}{\mathrm{argmin}}\left( \boxed{OF{_{S.R[j]}}(S.T[i])}\right)$
        S = $\bxt{assign(S, S.R[j], t)}$
\end{lstlisting}

\begin{lstlisting}[mathescape=true, caption={Declarative version of \texttt{Algorithm} in the SMT-LIB language.}, language=Python,belowskip=-8pt, label={lst:smtAlgorithm}]
(define-fun algorithm ((S state)) state 
    (ite empty((select (resources S) i)) S
            ($\bxt{assign}$ S (argmin (tasks S) $\bxt{OF})$ i)))
\end{lstlisting}
%\vspace{0.2in}
\end{figure}
\subsection{System}

The system is imperative code for a transition function $\texttt{System}$ which modifies a state variable and decides the time of the next invocation of the algorithm; in other words, it represents the passage of time and the completion of work in the given resource allocation problem. Intuitively, the system function checks whether any work is left. If there is none, it sets a ``done'' state and stops the running of the scheduling algorithm. Otherwise, it performs a problem-specific calculation to determine when the next allocation event should occur, updates the state of the relevant tasks and the resources they are using, and increments time and other system counters. This process is shown in function \texttt{System} of Listing~\ref{lst:imperativeSystem}.

The \texttt{System} function can represent any resource allocation problem with variation in three functions. In particular, users specify the three boxed functions in Listing~\ref{lst:imperativeSystem} describing a problem of interest. \texttt{doneCondition} is a predicate on state that returns true if all tasks in a system have finished or a global maximum running time has been reached. \texttt{nextEvent} is a function on state which determines the time of the next invocation of the scheduling algorithm. \texttt{update} is a function that modifies state for making progress on tasks. Listing~\ref{lst:smtSystem} gives an exact translation of \texttt{Algorithm} into a declarative function in the SMT-LIB language. \texttt{doneCondition}, \texttt{nextEvent}, and \texttt{update} are structurally declarative; each produces an output through functional computations on an input. Thus, they can either be plugged directly into the constraint in Listing~\ref{lst:smtSystem} or, if starting from an imperative definition, they can be translated into declarative constraints easily.

\subsection{Algorithm}

Invocations of the allocation algorithm punctuate the operation of the system; $\texttt{Algorithm}$ is also a transition function that modifies state. It represents the operation of the allocator by updating the state to reflect a change in resource allocation without the passage of any time. The imperative \texttt{Algorithm} function is shown in Listing~\ref{lst:imperativeAlgorithm}. The algorithm checks whether any resources are available in the current state. If they are, it finds the task best suited to that resource using a resource-specific objective function $\tt OF{_{S.R[j]}}$ and changes the state to reflect an assignment of this task to the given resources. Finally, it calls the system function.

In addition to the resource-wise objective function, users must provide a problem-specific definition of \texttt{assign}, which represents the dedication of a resource to performing work on a particular task. Assign may simply assign a binary variable for one task, or may represent a complex resource that performs different types of work on multiple tasks. Listing~\ref{lst:smtAlgorithm} shows an equivalent version of \texttt{Algorithm} in SMT-LIB format. Both \texttt{assign} and the objective function are declarative in nature and can be directly encoded as SMT constraints..

\subsection{Performance Queries} 

The above subsections give a precise conversion from imperative code representing a resource allocation problem into declarative constraints; along with the state description and state invariants, they allow an SMT solver to reason about a scheduling heuristic. We can now harness the power of the solvers to make performance queries. Queries can be of two types. First, we can encode queries that prove additional state invariants. Such queries can be directly expressed using the state SMT variables. We ask the solver to find a counterexample that satisfies all constraints of our model as well as the negation of the invariant. The property holds if the query is unsatisfiable. 

Our model can also be used to compare the performance of heuristics with {\em the} optimal algorithm for a specific metric such as the average task completion time. Thus, \framename can be used to derive upper bounds and worst-case scenarios for the performance of resource allocation heuristics.

Instead of describing the optimal algorithm explicitly, we use the power of the SMT solver to simulate it. First, we instantiate a copy of the state trace---this will be used to represent an ideal schedule. For comparison, the optimal state trace must have the same initial state as that subject to the scheduling algorithm. However, instead of the alternating calls to \texttt{System} and \texttt{Algorithm}, the ideal state trace is subject only to calls to \texttt{System}. This allows the solver to make any feasible ordering choices for the ideal state trace. Finally, we ask the solver to optimize the gap between the performance metric under the heuristic and under the optimal algorithm. For instance, we can ask the solver to find the worst average time to completion in a scheduling algorithm.

The rest of the paper demonstrates the generality and easy of use of \framename through four different case studies in which we draw actionable conclusions about the studied system. Appendix~\ref{sec:summary} summarizes how we fill the blanks in our framework for each case study. Further, appendices~\ref{sec:ccac} and~\ref{sec:fperf} show how we use our framework to replicate earlier modeling efforts by CCAC and FPerf.
%Next, we will go through multiple use cases of this framework, demonstrating its generality and ease of use. 
%\aditya{add a lead out. What are the next few sections about?}

%% file: 04_work_stealing.tex
\section{Case study: Work Stealing}
\label{sec:ws}
This section gives a motivating example applying the \framename framework to the problem of work stealing. Work stealing schedulers assign tasks to processors. Each processor executes tasks in its local queue in order, without preemption. When a processor becomes idle, and no task is available in its local queue, it steals the oldest task from another processor's queue. Work stealing is most commonly studied by arranging tasks as a Directed Acyclic Graph (DAG) with edges representing dependencies; a task can only be scheduled if all its dependencies have finished. A well-known theorem guarantees that work stealing will find a schedule for any DAG that finishes executing in at most twice the time of an offline optimal scheduler~\cite{ws}, but does not account for the cost of context switching between threads. We provide an example use of \framename to extend this traditional workload model to include switching costs.

\begin{figure*}[!t]
\label{fig:wsa}
\input{wsa}
\caption{Performance comparisons of work stealing heuristic to optimal}
\end{figure*}

\subsection{Model}
We develop a formal model for work stealing schedulers using the \framename framework. \statet consists of task-wise variables giving the current queue, position in that queue, total running time, time enqueued, switching cost, and a set of dependencies. Since any task can run on any processor, the topology of queues is just the complete graph with as many vertices as processors. In general, switching cost can depend on the tasks and the processor involved because of architectural features like NUMA, and the status of any caches. However, we adopt a simpler workload model, and assume that this cost is fixed for each resource.

Given a list of idle processors and their local queues, the work stealing algorithm provides a mapping from idle processors to ready tasks according to the rules described above. From these rules, we define the five functions required to make \algo and \sys. If a task's dependencies have finished, \algo's objective function with resource $R$, $\objfun _R$ produces the inverse queue position for tasks in R's queue (FIFO) or the inverse task enqueue time for tasks in other queues. Otherwise, it produces an infinite value, indicating that the task cannot yet be run. \assign increments the position of every task in the given queue, causing the selected task to run on the specified resource. The \sys function requires definitions of \doneCondition, \nextEvent, and \funUpdate. \doneCondition simply checks whether any tasks are still in a queue. \funUpdate updates all the other state variables, honoring the DAG edges and the switching costs. 

Listing~\ref{lst:ws-imp} show as an illustrative example the imperative definition of \nextEvent. This function decides the next time of scheduler invocation from the current state; its role is to find the earliest point in the future with idle processor(s) and available tasks. If a task is currently running, and all tasks on which it depends are not running (i.e., they have finished), then it reports the end time of the task as a possible next event. The function then returns the minimum such time over the set of tasks, returning the end of the system run as a default. In other words, the next event is simply the earliest time at which a running task finishes.

\begin{figure}[!t]

\begin{lstlisting}[mathescape=true, caption={Imperative definition of \texttt{nextEvent}.}, language=Python,belowskip=-8pt, label={lst:ws-imp}]
def nextEvent(S):
    end = $\underset{i \in {\tt S.T}}{\mathrm{argmin}}($S.T[i].start + S.T[i].length if 
        S.T[i].running and not 
        S.T[i].parent.running else S.end $)$
    return end
\end{lstlisting}
\begin{lstlisting}[mathescape=true, caption={Declarative version of \texttt{nextEvent} for work stealing in the SMT-LIB language; for clarity of notation, we define the per-task helper SMT function \texttt{findEnd} which is then passed as an argument to \texttt{argmin}.}, language=Python,belowskip=-8pt, label={lst:ws-dec}]
(define-fun findEnd ((S state) (T task)) Real 
    (ite (and T.running (not T.parent.running)) 
    (+ T.start T.length) S.end))

(define-fun nextEvent ((S state)) state 
    (argmin (tasks S) (findEnd S)))
\end{lstlisting}

\end{figure}

\begin{comment}
\begin{figure}[!t]
\begin{lstlisting}[caption={Declarative and imperative }, language=Python,belowskip=2pt, aboveskip=2pt, label={lst:ws}]
def System(state$_t$, state$_{t+1}$, work):
  # update time
  new_tasks = state$_t$.mapping.values()

def Algorithm(state):
  for p in state.free:
    cur_q = state.queues.find(p)
    if len(my_queue) > 0:
      state.mapping[p] = cur_q.pop_back()
    else:
      for q in state.queues:
        if q.proc in free and len(q) > 1:
          state.mapping[p] = q.pop_front()
          break
        if len(q) > 0:
          state.mapping[p] = q.pop_front()
          break

\end{lstlisting}
\vspace{-0.1in}
\end{figure}
%  #     according to the DAG and switching costs
\end{comment}

Given the model of the work stealing problem, we use \framename to create declarative SMT constraints; Listing~\ref{lst:ws-dec} gives an example for \texttt{nextEvent}. We model task lengths and the per-task switching cost each with a single real number. The DAG is encoded as a boolean adjacency matrix of size $N_T \times N_T$. Each state variable consists of lists of booleans or reals. For instance, {\tt free} is a list of booleans of size $N_R$. An entry in {\tt free} is set for {\tt state}$_t$ if and only if the processor corresponding to that index is idle at {\tt state}$_t$.{\tt time}. Similarly, the algorithm's output map is a boolean matrix of size $N_R \times N_T$. Since work stealing is work conserving, at least one task is scheduled at every invocation of the algorithm, so at most $N_T$ states are required to schedule any workload ($K = N_T$). The resulting constraints are reasonably simple; all constraints amount to $<500$ SLoC written using Z3's Python API~
\cite{z3}.

%Next, we model the state and workload as SMT variables. 
%The solver is allowed to arbitrarily decide the value of these cost variables. 

%With the state variables and the workload modeled, it is straightforward to transform the imperative code of \algo and \sys as SMT constraints. 

%
\subsection{Queries and Results}
\begin{figure}[!t]
\begin{subfigure}{0.45\textwidth}
    \centering
    \includegraphics[width=0.8\textwidth]{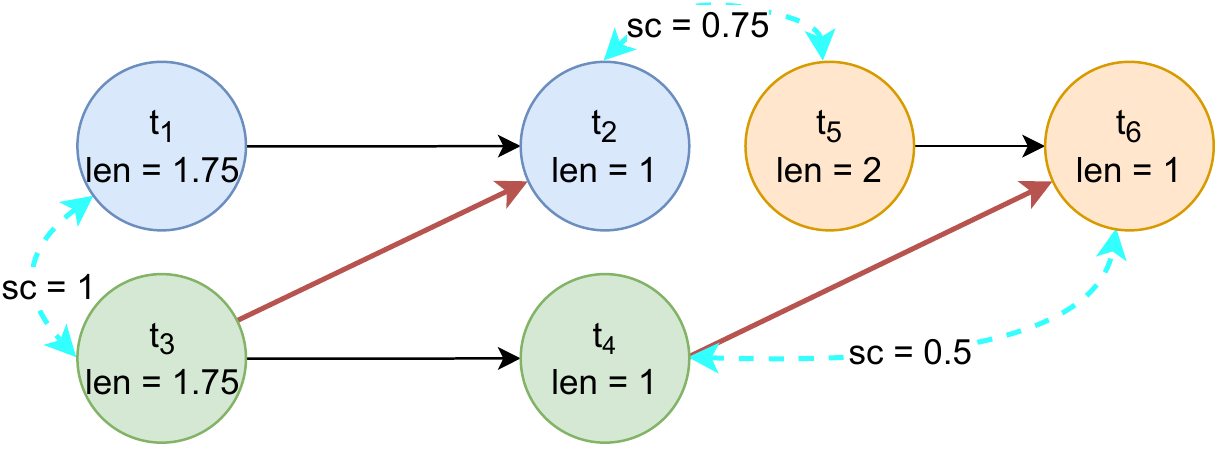}
    \caption{An example DAG. The color of each task represents the thread it belongs to. Red edges denote dependencies between tasks belonging to separate threads. The blue dashed lines show CPU jumps between tasks belonging to different threads, along with the switching cost (sc) incurred.}
    \label{fig:sample-dag}
\end{subfigure}
\begin{subfigure}{0.35\textwidth}
    % \centering
    \includegraphics[width=1.1\textwidth]{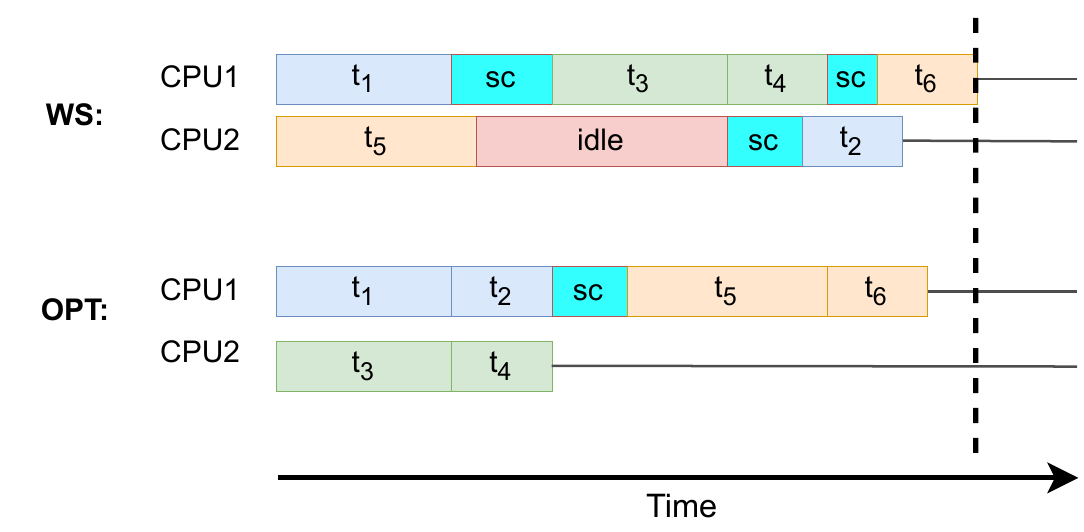}
    \caption{Work Stealing and Optimal schedules}
    \label{fig:sample-sched-ws}
\end{subfigure}
\vspace{0.05in}
\caption{Work Stealing vs Optimal}
\end{figure}

 To obtain performance results, we compare the total completion time of a work stealing scheduler with the optimal algorithm. by asking the solver to maximize the completion time ratio between work stealing and an optimal schedule.
 \begin{comment}
 \vspace{-0.15in}
\begin{align}
\texttt{maximize}(\texttt{time}_\texttt{ws} / \texttt{time}_\texttt{opt})
\end{align}
\end{comment}
Fig. \ref{fig:sample-sched-ws} shows an example work stealing schedule and compares it with the optimal schedule. %Our queries explore how this ratio varies as we impose different constraints on the switching costs.

First, we set the cost of switching to zero and query the maximal ratio between work stealing and optimal. We find that the bound is not $2$, but $2 - \frac{1}{N_R}$, where $N_R$ is the number of processors. Since we are using an SMT solver, the bound it found is exact and matches the known theoretical result \cite{graham}. Thus, even though we only queried for up to 4 processors and 9 tasks, we believe that the precision and consistency of our results allow for extrapolation to larger values.

Next, we introduce switching costs, which are parameterized in two ways: parameter $k$ caps the maximum switching cost and $c$ caps the difference between the switching costs of two tasks.
\begin{comment}
\begin{flalign}
&\forall i \in [N_T]\quad \texttt{switch\_cost}[i] \leq k \times \texttt{len}_{\texttt{min}}&\\
 &\max_{i \in N_T}(\texttt{switch\_cost}[i]) \leq c\times\min_{i \in N_T}(\texttt{switch\_cost}[i])&
\end{flalign}
\end{comment}
Since the unit of time is arbitrary, we may set maximum task length to $1$. We fix $c = 1$ and $k=10$. 
%This means that the solver can select any switching cost up to $10\times$ the task length, as long as the costs are equal to each other. 
Figure~\ref{fig:perf-with-dag} shows the optimality ratio as we vary the maximum size of the DAG tasks. For small DAGs, work stealing performs close to optimal, but performance worsens linearly as the size increases. The bound grows at a slower rate with an increasing number of CPUs and the ratio remains bounded even though switching cost can be up to $10\times$ the task length.

We also vary the bound on the maximum allowed switching cost while constraining all costs be equal to each other. Figure~\ref{fig:perf-with-k} plots the optimality ratio for different numbers of CPUs and tasks.\footnote{In realistic systems, $k$ should not be much larger than 1.} Interestingly, it plateaus to a value only slightly larger than in the case without switching cost. For instance, with 2 processors and up to 7 tasks, the optimality ratio is smaller than 4. Finally, we allow switching costs to vary between $0$ and $10$ as long as they are within a factor of $c$ of each other, as shown in Figure~\ref{fig:perf-with-c}. Here the ratio grows linearly with $c$, showing that the \emph{variation} in switching cost matters more than its relationship to task length.

%% file: wsa.tex
\begin{subfigure}{0.32\textwidth}
        \centering
        \begin{tikzpicture}[remember picture]
        \begin{axis}[%
        width=\textwidth,
        height=0.7\textwidth,
        y label style={at={(axis description cs:0.18,0.5)}},
        x label style={at={(axis description cs:0.5,0.05)}},
        xlabel={$N_T$},
        ylabel={WS / Optimal},
        scatter/classes={%
            a={mark=,draw=orange}}]
        \addplot[color=orange, mark=*]%
        table {
        x y
        1 1
        2 1
        3 1
        };
        \addplot[color=orange, mark=*]%
        table {
        x y
        3 1.8994140625
        4 2.7265625
        5 3.3330078125
        6 3.89404296875
        7 4.45
        };
        \addplot[color=blue, mark=diamond*]%
        table {
        x y
        1 1
        2 1
        3 1
        4 1
        };
        \addplot[color=blue, mark=diamond*]%
        table {
        x y
        4 2.63
        5 2.699951171875
        6 2.75
        7 2.81
        };
        \end{axis}
        \begin{scope}[scale=0.5, every node/.style={scale=0.5}]
        \matrix [draw] at (1.5,3.8) {
        %\matrix [draw, right=2pt] at (current bounding box.north east) {
            \node [draw,fill=orange,label=right:{$N_R=2$}] {}; \\
            \node [draw,fill=blue,label=right:{$N_R=3$}] {}; \\
        };
        \end{scope}
        \end{tikzpicture}
        \caption{Performance for different DAG sizes $(k = 10, c = 1)$. When $N_T < N_R$, work stealing is trivially optimal.}
        \label{fig:perf-with-dag}
\end{subfigure}
\hfill
\begin{subfigure}{0.32\textwidth}
            \centering
                \begin{tikzpicture}
                \begin{axis}[%
        width=\textwidth,
        height=0.7\textwidth,
        y label style={at={(axis description cs:0.18,0.5)}},
        x label style={at={(axis description cs:0.5,0.05)}},
        xlabel={$k$},
        ylabel={WS / Optimal},
        scatter/classes={%
            a={mark=,draw=orange}}]
        \addplot[color=blue, mark=.]%
        table {
        x y
        0 1.5
        0.01 1.6
        0.03 1.75
        0.05 1.875
        0.1 2.25
        0.2 2.75
        0.5 3.5
        1 3.75
        2 3.875
        5 3.9375
        10 3.96875
        };
        \addplot[color=orange, mark=.]%
        table {
        x y
        0 1.5
        0.01 1.5625
        0.03 1.6953125
        0.05 1.7998046875
        0.1 2.0
        0.2 2.3994140625
        0.5 3.0
        1 3.3330078125
        2 3.541015625
        5 3.7958984375
        10 3.8935546875
        };
        \addplot[color=green, mark=.]%
        table {
        x y
        0 1.666666
        0.01 1.69
        0.03 1.75
        0.05 1.9
        0.1 2.1
        0.2 2.39
        0.5 2.9
        1 3.25
        2 3.29
        5 3.32
        10 3.35
        };
        \addplot[color=red, mark=.]%
        table {
        x y
        0 1.666666
        0.01 1.6767578125
        0.03 1.6962890625
        0.05 1.7998046875
        0.1 2.0
        0.2 2.25
        0.5 2.5712890625
        1 2.7265625
        2 2.75
        5 2.75
        10 2.75
        };
        \end{axis}
        \begin{scope}[scale=0.5, every node/.style={scale=0.5}]
        \matrix [draw] at (5,1.1) {
        %\matrix [draw, right=2pt] at (current bounding box.north east) {
            \node [draw,fill=orange,label=right:{$N_R=2, N_T=6$}] {}; &
            \node [draw,fill=red,label=right:{$N_R=3, N_T=6$}] {}; \\
            \node [draw,fill=blue,label=right:{$N_R=2, N_T=7$}] {}; &
            \node [draw,fill=green,label=right:{$N_R=3, N_T=7$}] {}; \\
        };
        \end{scope}
        \end{tikzpicture}        
            \caption{Work Stealing performance as switching costs scale up. Every task has the same switching cost $(c = 1)$.}
            \label{fig:perf-with-k}
\end{subfigure}
\hfill
\begin{subfigure}{0.32\textwidth}
    \centering
        \begin{tikzpicture}
        \begin{axis}[%
        width=\textwidth,
        height=0.7\textwidth,
        y label style={at={(axis description cs:0.13,0.5)}},
        x label style={at={(axis description cs:0.5,0.05)}},
        xlabel={$N_T$},
        ylabel={WS / Optimal},
        scatter/classes={%
            a={mark=,draw=orange}}]
        \addplot[color=violet, mark=.]%
        table {
        x y
        1.1 2.8905410766601562
        1.2 3.054443359375
        1.5 3.5453948974609375
        2 4.363395690917969
        4 7.851417541503906
        8 14.620506286621094
        };
        \addplot[color=orange, mark=.]%
        table {
        x y
        1.1 3.5582351684570312
        1.2 3.7833175659179688
        1.5 4.4925537109375
        2 5.9049835205078125
        4 11.178558349609375
        8 20.19318389892578
        };
        \addplot[color=blue, mark=.]%
        table {
        x y
        1.1 4.1707916259765625
        1.2 4.446479797363281
        1.5 5.259193420410156
        2 6.573432922363281
        4 11.378715515136719
        8 20.39318389892578
        };
        \addplot[color=brown, mark=.]%
        table {
        x y
        1.1 2.779510498046875
        1.2 2.8595733642578125
        1.5 3.099761962890625
        2 3.4993209838867188
        4 5.099822998046875
        8 8.29931640625
        };
        \addplot[color=red, mark=.]%
        table {
        x y
        1.1 2.9199981689453125
        1.2 3.0899429321289062
        1.5 3.5997772216796875
        2 4.449501037597656
        4 7.892204284667969
        8 14.620506286621094
        };
        \addplot[color=green, mark=.]%
        table {
        x y
        1.1 3.513671875
        1.2 3.70703125
        1.5 4.48046875
        2 5.9049835205078125
        4 11.178558349609375
        8 20.19318389892578
        };
        \end{axis}
        \begin{scope}[scale=0.3, every node/.style={scale=0.5}]
        \matrix [draw] at (2.9,5.6) {
        %\matrix [draw, right=2pt] at (current bounding box.north east) {
            \node [draw,fill=violet,label=right:{$2,4$}] {}; 
            &
            \node [draw,fill=brown,label=right:{$3,5$}] {}; \\
            \node [draw,fill=orange,label=right:{$2,5$}] {}; 
            &
            \node [draw,fill=red,label=right:{$3,6$}] {}; \\
            \node [draw,fill=blue,label=right:{$2,6$}] {};
            &
            \node [draw,fill=green,label=right:{$3,7$}] {}; \\
        };
        \end{scope}
        \end{tikzpicture}
    \caption{Work Stealing performance when costs between different pairs of tasks can vary. $k = 10$, legend labels are $N_R, N_T$.}
    \label{fig:perf-with-c}
\end{subfigure}

%% file: 05_linux.tex
\section{Case study: The Linux CFS Load Balancer}
\label{sec:linux}

The Linux CFS load balancer underwent a major rework in 2019. We started our modeling effort in 2022, focusing on Linux v5.5, the version resulting from the rework. Our objective was to identify two specific types of performance bugs: wasted work (i.e., violation of work conservation) and unfairness (i.e., tasks receiving less than their fair share of CPU time due to load balancing decisions). After identifying one bug of each type (\S\ref{sec:linux-5.5}), we found that they were resolved in v5.7. However, we found new bugs of both types in v5.7 (\S\ref{sec:linux-5.7}). Instead of tracing the evolution of the code base, we model the latest kernel version v6.8-rc1, showing that some performance bugs persisted from v5.7 while new ones arose (\S\ref{sec:linux-6.8}). Adapting our model to various versions required less than a single day of work each.

\subsection{Model}

\noindent \textbf{Heuristic description.} The load balancer is optimized for multi-core architectures, capturing the proximity between cores by dividing them into a hierarchy of scheduling domains based on the SMT, SMP, and NUMA groups to which they belong. In particular, the load balancer preserves locality by first trying to balance tasks among SMT cores, then among SMP cores, then finally among NUMA nodes. When all CPUs are busy, balancing is done at regular intervals, with each CPU traversing its domain hierarchy, performing load balancing at each level. To avoid having all CPUs load balancing at all levels, CPUs are grouped within each scheduling domain, with a single CPU responsible for balancing for that group at that a given domain level. An example of a domain hierarchy of four CPUs is shown in Figure~\ref{fig:lb-topology}. At the top level, only one CPU from each group can move tasks between the two groups. If there is considerable imbalance between the two groups, the balancer picks the busiest CPU in the busiest group to steal tasks from. It tries to steal as many tasks as would relieve the source group of excess load and/or fairly distribute the load between the source and destination groups. Linux supports multiple task types, including CFS tasks, and the higher-priority realtime tasks. The load balancer only can only migrate CFS tasks. 

\noindent \textbf{Modeling scope.} The load balancer has various optimization objectives including minimizing imbalance, fairness, work conservation, and preserving locality. We limit our scope to verifying fairness and work conservation whose primary cause is load balancing behavior. For example, we assume that the CFS is completely fair, ignoring unfairness caused by CFS between tasks assigned to the same CPU. While uncovering unfairness issues in the CFS scheduler is interesting in its own right, it's out of scope.

We assume that the load balancer has a fixed number of compute-intensive CFS tasks that are all of the same priority, with none of them pinned to a specific CPU. These assumptions provide the load balancer with the most flexibility in moving tasks around. With a workload made primarily of of CPU-intensive tasks, we avoid modeling the behavior of the load balancer triggered when a CPU becomes newly idle.  
Due to our focus on work conservation and fairness, we relax all the constraints imposed on the algorithm when migrating tasks to maintain locality, providing it with freedom to minimize imbalance and avoid wasting CPU time. %Finally, we focus on the periodic behavior of the load balancer by assuming that all tasks are CPU intensive.\footnote{This assumption allows us to avoid modeling the behavior of the load balancer triggered when a CPU becomes newly idle by ensuring that all CPUs are either always busy or can't steal any work.} %We say that one timestep has passed after balancing the top-level domain once.

%\subsection{From Imperative to Declarative}
\noindent\textbf{Model detail.} The state of the Linux model has task--wise variables indicating which CPU's queue a task belongs to, the percentage of time the task has run recently, and the percentage it has been runnable. Resource--wise variables represent the domain hierarchy of CPUs; each processor has a list of parent domains. With this description of state, we next provide the functions necessary to make \algo and \sys work. 

The objective function of \algo is complex for the Linux scheduler, and based on the Linux \texttt{load\_balance()} function. We provide imperative pseudocode for this function in Appendix~\ref{alg:linux}. Listing~\ref{lst:linux-imperative} shows a small representative sample in imperative code. The subset of the objective function is based on the DetachTasks function, which is responsible for picking and migrating tasks between a busy CPU and a load balancing CPU. In this example, the objective function returns either $1$, indicating that a task cannot migrate, or $0$, indicating that it can. Note that these values are further modified by other parts of the objective function ommitted from Listing~\ref{lst:linux-imperative}. Variable \texttt{num\_migr} represent the number of tasks migrating at the current event and \texttt{sum\_migr} represents the sum of metrics of these migrating tasks. Listing~\ref{lst:linux-declarative} gives the declarative version of the objective function excerpt; by using \framename, we can see that the translation from Listing~\ref{lst:linux-imperative} to Listing~\ref{lst:linux-declarative} is nearly trivial.

In contrast to \algo, \sys for the Linux load balancer is simple, highlighting the simplifying power of \framename's overapproximation approach. There is no \doneCondition; models can progress until the maximum number of steps is reached. \nextEvent is also simple since the balancer is invoked periodically; therefore \nextEvent is just a fixed increment from the previous event. Finally, \funUpdate increments the progress of each task by its share of running time among tasks in its current queue.

\begin{figure}[!t]
\begin{lstlisting}[caption={Imperative code for an excerpt of the objective function.}, language=Python,belowskip=2pt, aboveskip=2pt, label={lst:linux-imperative}]
def $\texttt{OF}_{R}(\texttt{T})$:
    if sum_migr >= R.imbalance or
        T.metric / 2 > (R.imbalance - sum_migr) or
        num_migr + 1 == R.num_running:
        #Cannot migrate
        return 1
    else:
    #Can migrate (T.metric / 2 <= R.imbalance)
        return 0
\end{lstlisting}
\vspace{-0.1in}
\begin{lstlisting}[caption={Declarative version of the impeative code in Listing~\ref{lst:linux-imperative}.}, language=Python,belowskip=2pt, aboveskip=2pt, label={lst:linux-declarative}]

(define-fun OF ((R resource) (S state)) Real 
    (ite (or (>= sum_migr R.imbalance) 
        (> (/ T.metric 2) (- R.imbalance sum_migr)) 
        (= (+ num_migr 1) R.num_running)) 0 1))
\end{lstlisting}
\vspace{-0.2in}
\end{figure}

\subsection{Queries and Results} \label{sec:linux-results}

% For the topology in \jehad{can't we add the diagram that shows how the topolgy is organized into groups and domains?}, the lower domains are balanced twice as frequently as the higher domain. In our model, we say that one timestep has passed after balancing the top-level domain once.

% We formulate two kinds of queries: one to check work conservation and another to check fairness. For work conservation, we run the model with one timestep and ask if a CPU which starts idle can end up idle. For fairness, we run it for up to 4 timesteps and ask if, at the end of the trace, one task can get less than some fraction of the execution time that some other task gets.

For the topology shown in Figure~\ref{fig:lb-topology}, we ran our queries to check work conservation and fairness. In particular, we consider the load balancer non-work conserving, if a CPU is idle after the load balancer is invoked when there are more tasks than there are CPUs; and we consider it unfair, if the ratio between the task receiving the least CPU time and the task receiving the most CPU time is below a certain threshold (e.g., 0.4).

\subsubsection{Linux v5.5\\}

\label{sec:linux-5.5}
%\jehad{why does this render so terribly?}

\noindent \textbf{Work Conserving.} As shown in Appendix~\ref{alg:linux}, before migrating a task, the load balancer has to determine the basis of the migration, referred to herein as the migration type. There are multiple migration types. The type \texttt{MIGRATE\_UTIL} is chosen when the current CPU's group has spare capacity, and another group is overloaded. The latter is deemed the busiest group.
Such a decision can be made in the scenario shown in Figure~\ref{fig:lb-ex} when the utilization average of all tasks in Group 1 is high, and Group 2 only has a single task.

The algorithm chooses \texttt{MIGRATE\_UTIL} to balance the utilization average amongst groups by stealing tasks from the busiest CPU of the busiest group. The CPU with the highest utilization average in the busiest group is determined to be the busiest. A CPU's utilization average is just the sum of utilization averages of the tasks in its queue. The utilization average of a task is defined as the weighted moving average of its running time. 

In our example, CPU2 can be the busiest CPU if its only task has a higher utilization average than the sum of the utilization averages of tasks on CPU1. However, CPU3 cannot steal work from CPU2 since it only has a single runnable task. This constraint helps avoid bouncing tasks between idle CPUs, but causes the scheduler to not be work conserving. One way to fix this bug is to define the busiest CPU to be the one with the highest utilization average that has more than one runnable task. %(line~\ref{l:added-in-5.7}). 

\noindent \textbf{Fairness.} We ask whether a task can get less than 40\% the execution time of another task. One trace in which this happens starts with the work conserving bug described above. Under our assumption that tasks are CPU-bound, it takes a few timesteps for the utilization of CPU1 to increase as the running time of its tasks increases, after which the task distribution starts to converge to an even one. However, during that time, tasks are treated unfairly. Even worse, as soon as the load becomes even (after 4 invocations of top-level domain balancer correspond to approximately 16 jiffies or 160ms), it is plausible that the three tasks on CPU1 would block, thus lowering their stats, and the sequence repeats again.

After finding the work conserving bug and assessing its impact on fairness using this model, we realized that it was also identified by the Linux community and fixed in Linux v5.7~\cite{b1-commit}. %We then moved to v5.7.

\subsubsection{Linux v5.7\\}\label{sec:linux-5.7}

\noindent \textbf{Work Conserving.} When the busiest group is not overloaded, the load balancer applies a less aggressive migration type. In that scenario, if the group performing load balancing has a single idle CPU, that CPU remains idle because the less aggressive migration type only migrates tasks if the difference in the number of idle CPUs between the two groups is larger than one. Arguably, this bug is not detrimental to performance as it only occurs when none of CPU groups are overloaded.

\begin{figure}[!t]
\centering
\begin{subfigure}{0.28\textwidth}
    \centering
    \includegraphics[width=1\textwidth]{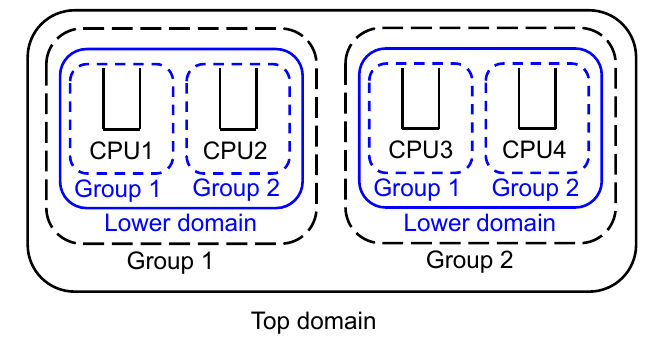}
    \vspace{-0.2in}
    \caption{Topology}
    \label{fig:lb-topology}
\end{subfigure}
\begin{subfigure}{0.19\textwidth}
    \centering
    \includegraphics[width=1\textwidth]{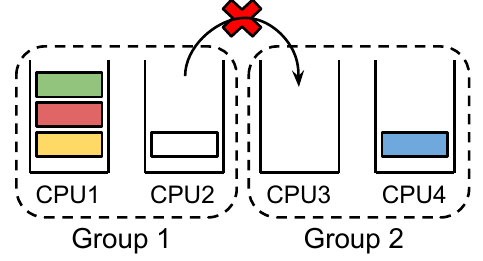}
    \vspace{-0.2in}
    \caption{Example of the work conservation bug in v5.5}
    \label{fig:lb-ex}
\end{subfigure}
\caption{The \texttt{runqueue}s of each CPU. Tasks are represented as rectangles.}
\label{fig:lb-ex33}
\end{figure}

\noindent \textbf{Fairness.} We found that v5.7 can lead to unfairness when the imbalance between groups of CPUs is smaller than the minimum imbalance needed to make the migration decision. For example, in the top domain in Figure~\ref{fig:lb-topology}, if one group has three tasks while the other has only two, the imbalance between the two groups is deemed too small. Yet, we found that this bug was fixed by the Linux community in v5.10~\cite{fairness-commit}. The fix essentially introduces a counter that gets incremented when an imbalance is detected yet no task is stolen, leading the load balancer to be aggressive the next round, lowering its imbalance threshold.

\subsubsection{Linux v6.8\\}\label{sec:linux-6.8}

\noindent \textbf{Work Conserving.} The bug we found in v5.7 persisted.

\begin{figure}[!t]
    \centering
    \includegraphics[width=1\linewidth]{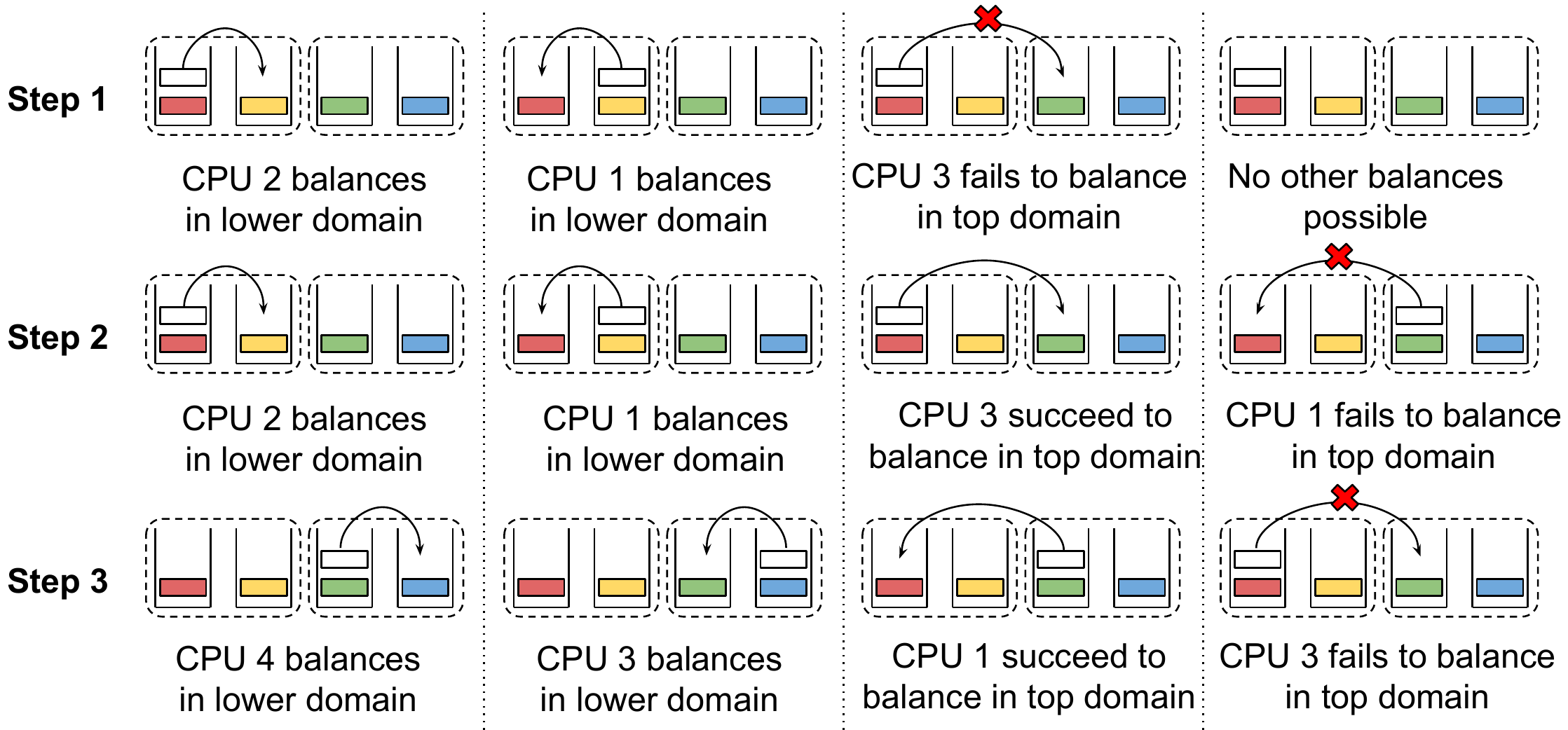}
    \caption{The task, represented by the white box, receives less than 60\% of the CPU time of the yellow and blue tasks. Every failed balance increments the \texttt{nr\_balance\_failed}, making balancing more aggressive during the following event. The domain hierarchy and all CPU and group labels are similar to that of Figure~\ref{fig:lb-topology}. }
    %\ahmed{a more descriptive caption is needed as well as a labeling of the Xs in the figure.}\sg{Perhaps the Xs can be labelled with the Bug they refer to?}} \ahmed{yes, that's what I meant}
    \label{fig:lb-ex2}
\end{figure}

\noindent \textbf{Fairness.} We asked if one task can get less than 60\% the execution time of another task. The trace produced shows that tasks migrate across groups as intended. Specifically, after a failed migration attempt, the load balancer becomes more aggressive. However, since load balancing can be a executed simultaneously by multiple CPUs, a sequence of events, shown in Figure~\ref{fig:lb-ex2}, can happen in which a task only bounces between CPU1 and CPU3. 
%Although, this is one particularly unlucky sequence of events, but the same effect can be produced by a large number of its permutations, so this type of unfairness is not improbable. 
The reason for this bug is that load balancing is synchronous but uncoordinated, and it also does not track the migration history of tasks.%; on the contrary, the only history tracked is cache-hotness, which discourages task migration.

% \begin{figure}[!t]
% \centering
% \includegraphics[width=1\linewidth]{figures/topology.pdf}
% \vspace{-0.25in}
% \caption{The \texttt{runqueue}s of each CPU. Tasks are represented as rectangles.}
% \label{fig:lb-ex}
% \end{figure}

It is worth noting that both these issues intricately depend on the workload characteristics generated by the solver. Slight deviation can lead to a completely different outcome. For instance, a different migration type may apply as the number of tasks or the blocking pattern of existing tasks evolves. A different measure of imbalance may be able to push tasks to idle CPUs and balance load more evenly in general.  Finely controlling the workload characteristics in synthetic benchmarks that highlight these bugs is difficult. This makes it hard for fuzzers and other tests to detect them. However, it does not preclude them from appearing in the real world. Our framework does not face these limits, and can freely search the workload space to generate intricate violating traces. 
%\vspace{-0.1in}
\begin{comment}
\begin{center}
\noindent\fbox{%
    \parbox{.9\linewidth}{
        \framename allows developers to find and fix bugs in real large systems. We discovered two bugs in the Linux CFS load balancer which result in wasted CPU time.
    }}
\end{center}
\end{comment}

%% file: appendix.tex
\begin{appendices}
\section{Summary of Case Studies}
\label{sec:summary}

Table~\ref{tab:summary} provides a summary of all the cases studies covered in this paper. The table presents all the details of how to fill the blanks in listings 1--4.  

\begin{table*}[]
\centering 
\scriptsize
\begin{tabular}{l|lll|ll|lll|}
\cline{2-9}
                                                                               & \multicolumn{3}{c|}{\textbf{State}}                                                                                                                                                                                                                                                                       & \multicolumn{2}{c|}{\textbf{Algorithm}}                                                                                                                                                                        & \multicolumn{3}{c|}{\textbf{System}}                                                                                                                                                                                                                                                                                                  \\ \cline{2-9} 
                                                                               & \multicolumn{2}{c|}{\textbf{Tasks}}                                                                                                                                                                              & \multirow{2}{*}{\textbf{\begin{tabular}[c]{@{}l@{}}Resource \\ Topology\end{tabular}}} & \multicolumn{1}{l|}{\multirow{2}{*}{\textbf{Objective Func.}}}                                                       & \multirow{2}{*}{\textbf{Assign}}                                                        & \multicolumn{1}{l|}{\multirow{2}{*}{\textbf{doneCondition}}}                                            & \multicolumn{1}{l|}{\multirow{2}{*}{\textbf{nextEvent}}}                                                                & \multirow{2}{*}{\textbf{Update}}                                                                  \\ \cline{2-3}
                                                                               & \multicolumn{1}{l|}{\textbf{Bookkeeping}}                                                                                         & \multicolumn{1}{l|}{\textbf{Deps.}}                                          &                                                                                        & \multicolumn{1}{l|}{}                                                                                                &                                                                                         & \multicolumn{1}{l|}{}                                                                                   & \multicolumn{1}{l|}{}                                                                                                   &                                                                                                   \\ \hline
\multicolumn{1}{|l|}{\begin{tabular}[c]{@{}l@{}}Work \\ Stealing\end{tabular}} & \multicolumn{1}{l|}{\begin{tabular}[c]{@{}l@{}}Running time\\ Current queue\\ Position in the queue\\ Time enqueued\end{tabular}} & \multicolumn{1}{l|}{\begin{tabular}[c]{@{}l@{}}Arbitrary\\ DAG\end{tabular}} & \begin{tabular}[c]{@{}l@{}}Complete \\ graph\end{tabular}                              & \multicolumn{1}{l|}{\begin{tabular}[c]{@{}l@{}}FIFO if work \\ available and steal \\ oldest otherwise\end{tabular}} & \begin{tabular}[c]{@{}l@{}}Change position\\ in queue\end{tabular}                      & \multicolumn{1}{l|}{\begin{tabular}[c]{@{}l@{}}All tasks \\ finished\end{tabular}}                      & \multicolumn{1}{l|}{\begin{tabular}[c]{@{}l@{}}A resource \\ becomes\\ available\end{tabular}}                          & \begin{tabular}[c]{@{}l@{}}Change current \\ queue and position \\ in queue\end{tabular}          \\ \hline
\multicolumn{1}{|l|}{SRPTF}                                                    & \multicolumn{1}{l|}{\begin{tabular}[c]{@{}l@{}}Running, \\ Blocking times\\ Current stage\end{tabular}}                           & \multicolumn{1}{l|}{None}                                                    & 1 queue                                                                                & \multicolumn{1}{l|}{\begin{tabular}[c]{@{}l@{}}Remaining \\ running time\end{tabular}}                               & \begin{tabular}[c]{@{}l@{}}Change current\\ stage to running\end{tabular}               & \multicolumn{1}{l|}{\begin{tabular}[c]{@{}l@{}}All tasks in\\ finished stage\end{tabular}}              & \multicolumn{1}{l|}{\begin{tabular}[c]{@{}l@{}}A task becomes\\ available to run\end{tabular}}                          & \begin{tabular}[c]{@{}l@{}}Change stage\\ of task\end{tabular}                                    \\ \hline
\multicolumn{1}{|l|}{\begin{tabular}[c]{@{}l@{}}TCP\\ Sync.\end{tabular}}      & \multicolumn{1}{l|}{\begin{tabular}[c]{@{}l@{}}\#Servers x \\ (sum, bp, bc,\\ ready to send, \\ \#rounds)\end{tabular}}           & \multicolumn{1}{l|}{None}                                                    & \begin{tabular}[c]{@{}l@{}}Arbitrary \\ graph\end{tabular}                             & \multicolumn{1}{l|}{\begin{tabular}[c]{@{}l@{}}All tasks ready \\ to send\end{tabular}}                              & \begin{tabular}[c]{@{}l@{}}Change counters\\ based on allocated\\ capacity\end{tabular} & \multicolumn{1}{l|}{None}                                                                               & \multicolumn{1}{l|}{\begin{tabular}[c]{@{}l@{}}A server finishes\\ computing or a flow\\ finishes sending\end{tabular}} & \begin{tabular}[c]{@{}l@{}}Change bookkeeping\\ variables to reflect\\ progress\end{tabular}      \\ \hline
\multicolumn{1}{|l|}{Linux}                                                    & \multicolumn{1}{l|}{\begin{tabular}[c]{@{}l@{}}Running \\ Runnable\\ Current queue\end{tabular}}                                  & \multicolumn{1}{l|}{None}                                                    & \begin{tabular}[c]{@{}l@{}}Sched. \\ domain \\ hierarchy\end{tabular}                  & \multicolumn{1}{l|}{Rebalance load}                                                                                  & \begin{tabular}[c]{@{}l@{}}Change current\\ queue\end{tabular}                          & \multicolumn{1}{l|}{None}                                                                               & \multicolumn{1}{l|}{Periodic}                                                                                           & \begin{tabular}[c]{@{}l@{}}Change running \\ based on capacity\end{tabular}                       \\ \hline
\multicolumn{1}{|l|}{CCAC}                                                     & \multicolumn{1}{l|}{\begin{tabular}[c]{@{}l@{}}cwnd\\ acks\end{tabular}}                                                          & \multicolumn{1}{l|}{None}                                                    & 1 queue                                                                                & \multicolumn{1}{l|}{N/A}                                                                                             & CCA (e.g., AIMD)                                                                        & \multicolumn{1}{l|}{None}                                                                               & \multicolumn{1}{l|}{\begin{tabular}[c]{@{}l@{}}Two consecutive\\ events separated\\ by an RTT\end{tabular}}             & Update \#acks                                                                                     \\ \hline
\multicolumn{1}{|l|}{FPerf}                                                    & \multicolumn{1}{l|}{\begin{tabular}[c]{@{}l@{}}Time enqueued\\ Current queue\\ Priority\end{tabular}}                             & \multicolumn{1}{l|}{None}                                                    & \begin{tabular}[c]{@{}l@{}}Arbitrary\\ DAG\end{tabular}                                & \multicolumn{1}{l|}{\begin{tabular}[c]{@{}l@{}}FIFO \& Priority\\ queueing\end{tabular}}                             & \begin{tabular}[c]{@{}l@{}}Changes current\\ queue of task\end{tabular}                 & \multicolumn{1}{l|}{\begin{tabular}[c]{@{}l@{}}All tasks are\\ dequeued from\\ all queues\end{tabular}} & \multicolumn{1}{l|}{\begin{tabular}[c]{@{}l@{}}Next dequeue time \\ based on rates of \\ all resources\end{tabular}}    & \begin{tabular}[c]{@{}l@{}}Assign enqueue time\\ and move tasks \\ between resources\end{tabular} \\ \hline
\end{tabular}
\vspace{0.1in}
\caption{A summary of the encoding of all six case studies, covered in the paper.}
\label{tab:summary}
\end{table*}

\section{CCAC}
\label{sec:ccac}

CCAC is a tool that uses an SMT solver, Z3~\cite{z3}, to verify the performance properties of congestion control algorithms. To test the generality of \framename, we re-implemented CCAC. This was straightforward since CCAC already maintains separate state variables for each time-step and separates the network model from the algorithm. One limitation in CCAC's original encoding is that computing the RTT requires computing the intersection between two lines. This needs a constraint involving a real-valued division between two solver-chosen variables. Z3 cannot handle the resulting non-linearity and times out without producing an output. To circumvent this problem, CCAC over-approximates the result as shown in Figure 5B of the paper~\cite{ccac}, leading to looser bounds than necessary.

More specifically, CCAC maintains $A[t]$ and $S[t]$ as arrays of real variables that denote the cumulative number of bytes that have arrived into and have been served from the network respectively. To compute the RTT at time $t$, we need to find $t' < t$ such that $A[t'] = S[t]$. Now, $A$ and $S$ are only defined at integer time points. Thus we must linearly interpolate $A$. To do so, we first find a $t_0 \in \mathbb{N}$ such that $A[t_0] \le S[t] \wedge S[t_0 + 1] > S[t]$. We know that $t' \in [t_0, t_0 + 1)$. Doing the calculation, we get $t' = t_0 + \frac{S[t] - A[t_0]}{A[t_0 + 1] - A[t_0]}$. The RTT equals $t - t'$ and clearly involves a division between SMT variables.

\framename on the other hand advocates for letting the solver chose when events occur by allowing the solver to pick the variables that the determine the output of \nextEvent. Thus, we can insist that, for every $t$, the solver also include an event at $t'$, except in the base case where the resulting $t'$ would be negative. This way $t$ and $t'$ are both SMT-chosen variables and $\mathrm{RTT} = t - t'$ can be computed using purely linear constraints. For a similar reason, we can avoid the over-approximation that CCAC made in Figure 5A of the paper.

More precisely, we define \nextEvent at a time $t$ as a non-deterministic function that allows for all times between $t$ and the minimum among $t_x = t' + D$ and $t_y = t' + \mathrm{RTT}$ among all $t' < t$ such that $t_x > t$ or $t_y > t$ respectively.

\section{Case study: Packet Scheduling}
\label{sec:fperf}
\subsection{Overview}
FPerf \cite{fperf} is a framework that models the flow of packets through sequences of queuing modules. It can answer queries about individual packets as well as the aggregate metrics by generating workloads that satisfy the query, if any. A workload, as FPerf defines it, is the rate of packet arrival at input queues.

In addressing packet scheduling, we demonstrate how our methodology facilitates and organizes thinking about modelling heuristics. It reduces the need for a full-fledged application-specific framework because the models produced are expressive and easily extensible.
\subsection{Model}
In this section, we call tasks packets because they have uniform processing times. The distinctive feature of queuing modules is the FIFO queue, which holds packets ordered by the time of enqueuing. \algo invocations are thus defined in terms of dequeuing events. Between two invocations, \sys is free to introduce up to K new packets in the module, where K is specified by the user. State invariants include that a packet is dropped iff it arrives at a full queue, and that new packets only arrive at input queues. Hence we can conclude that state variables must express the arrival time of each packet in a queue, from which the size of the queue can be derived.

We start with the simplest queueing module, namely a single input queue and a single output queue. \sys and \algo are one liners: \sys assigns unique timestamps to arriving packets and \algo picks the packet with the smallest timestamp to transition to the output queue.

Moving one step further, we model a priority scheduler module, with N input queues and one output queue. The only change in \sys is to repeat the creation of packets N times. In contrast, now \algo must pick the highest priority non-empty queue, in addition to picking the earliest arriving packets in the chosen queue.

On the same level of complexity is the round-robin scheduler. Whereas \sys stays the same, \algo needs to keep track of how many times a queue is polled. This warrants an extension in \state variables: increment the poll tally in the next state for a queue if it is dequeued in the current state or should have been chosen but was found empty.

To model FQ-CoDel, \state will be updated to capture the new classification of queues (\textit{new\_queues}, \textit{old\_queues} and \textit{inactive} queues), and \algo will extend accordingly.

Moving even more steps further, to model a longer sequence of queues, the state invariants will include the topology constraints. \algo will not only pick queues and packets to dequeue, but will also pick next queues in which to move the dequeued packets. \sys all the while stays the same.

\subsection{Queries and Results}
We queried the priority scheduler to find if the third highest-priority queue can be blocked for more than 5 invocations of \algo, and it produced a trace in which the highest and second highest priority queues were served and starved the third.

While neither is this surprising, nor is a single trace as helpful as an entire workload, the point here is the ease with which developers can create a model when thinking in terms of our methodology. Given a suitable definition of "workload", workload synthesis can always be added on top of the model.

\newpage
\section{Linux Load Balancer pseudo-code}

Snippet of pseudo-code of the {\tt load\_balance()} and related function in Linux v5.5 

{\footnotesize

% New definitions
\algnewcommand\algorithmicswitch{\textbf{switch}}
\algnewcommand\algorithmiccase{\textbf{case}}
\algnewcommand{\LeftComment}[1]{\Statex \(\triangleright\) #1}
\algnewcommand\procspace{\vspace{5pt}}
% New "environments"
\algdef{SE}[SWITCH]{Switch}{EndSwitch}[1]{\algorithmicswitch\ #1\ \algorithmicdo}{\algorithmicend\ \algorithmicswitch}%
\algdef{SE}[CASE]{Case}{EndCase}[1]{\algorithmiccase\ #1}{\algorithmicend\ \algorithmiccase}%
\algtext*{EndSwitch}%
\algtext*{EndCase}%

\algtext*{EndFor}% Remove "end for" text
\algtext*{EndIf}% Remove "end if" text

\begin{algorithm}
\caption{Linux CFS Load Balancing Algorithm}\label{alg:linux}
\begin{algorithmic}[1]

\Procedure{LoadBalance}{$c, sd$}
\If{$!\textrm{ShouldWeBalance}(c, sd)$} \label{l:isresponsible}
    \State \Return
\EndIf
\If{$!\textrm{ConsiderableImb}(c.idle, sd)$} \label{l:considerableimb}
    \State \Return
\EndIf
\State $dst\_g \gets sd.\textrm{FindGroup}(c)$
\If{$!sd.\textrm{GroupAboveAvg}(dst\_g)$}
    \State \Return
\EndIf
\State $src\_g \gets sd.\textrm{FindBusiestGroup}()$\label{l:findbusiestg}
\State $m\_type \gets \textrm{MigrationType}(src\_g, dst\_g, c.idle)$\label{l:migtype}
\State $src\_c \gets \textrm{BusiestCPU}(src\_g.CPUs, m\_type)$
\State $imb \gets \textrm{CalcImb}(src\_g, dst\_g, m\_type)$
\State $\textrm{DetachTasks}(src\_c, c, m\_type, imb)$ \\ \Comment{moves tasks based on migration type}
\EndProcedure

% \Require $n \geq 0$
% \Ensure $y = x^n$
\Procedure{MigrationType}{$busiest, local, idle$}
  \If{$local.\textrm{HasSpare}()$}\ 
    \If{$busiest.\textrm{Overloaded}()$}
        \State $has\_cap \gets local.capacity > local.util$
        \If{$! idle \OR has\_cap$}
            \State \Return{\texttt{MIGRATE\_UTIL}}\label{l:migutil}
        \Else \
            \State \Return{\texttt{MIGRATE\_TASK}}
        \EndIf
    \EndIf
    \State $\cdots$
    % \Else \ $\cdots$
  \EndIf
\EndProcedure
\procspace
\Procedure{BusiestCPU}{$CPUs,m\_type$}
    \State $key1 \gets \textbf{lambda}\ c : c.util\_avg$
    \State $key2 \gets \textbf{lambda}\ c : c.nr\_running$
    \State $moreThanOne \gets \textrm{Filter}(CPUs, key2)$ \\ \Comment{added in v5.7}\label{l:added-in-5.7}
    \Switch{$m\_type$}
        \Case{\texttt{MIGRATE\_UTIL}}
            \State \Return $\textrm{argmax}(moreThanOne,\  key1)$
        \EndCase
        \Case{\texttt{MIGRATE\_TASK}}
            \State \Return $\textrm{argmax}(CPUs,\  key2)$
        \EndCase
        \Case{$\cdots$} \EndCase
    \EndSwitch
\EndProcedure

%\procspace

% \algstore{lb}
% \end{algorithmic}
% \end{algorithm}

% \begin{algorithm}
%     \begin{algorithmic}[1]
%\algrestore{lb}

\Procedure{DetachTasks}{$src, dst, m\_type, imb$}
     \For{$task \textbf{ in } src.tasks$}
        \State $val \gets task.\textrm{Metric}(m\_type)$
        \If{$val \leq 2 * imb \AAND \textrm{CanMigrate}(src, task)$}
        \\ \Comment{changed after v5.7}\label{l:added-in-5.7}
            \State $\textrm{migrate}(task, src, dst)$
            \State $\texttt{imb} \gets \texttt{imb} - \texttt{val}$
        \EndIf
      \EndFor
\EndProcedure
\procspace

\Procedure{CalcImb}{$busiest, local, m\_type$}
    \Switch{$m\_type$}
        \Case{\texttt{MIGRATE\_UTIL}}
            \State \Return $local.capacity - local.util$
        \EndCase
        \Case{\texttt{MIGRATE\_TASK}}
            \If{busiest.\textrm{Overloaded}()}
                \State \Return $1$
            \Else
                \State $t_1 \gets busiest.nr\_idle$
                \State $t_2 \gets local.nr\_idle$
                \State \Return $\max(0, (t_1 - t_2) / 2)$ \label{l:task-imb}
            \EndIf
        \EndCase
        \Case{$\cdots$} \EndCase
    \EndSwitch
\EndProcedure

 \end{algorithmic}
\end{algorithm}
}

\section{SRPT Example}\label{app:srpt}
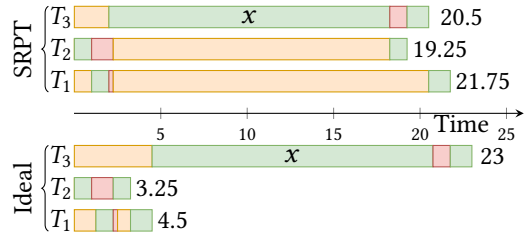
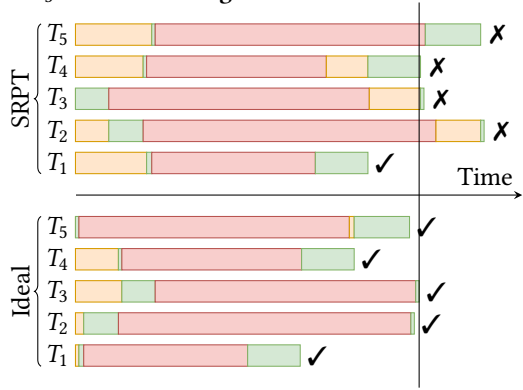
\begin{figure}[!t]
\centering
\begin{subfigure}{\linewidth}
    \centering
    \pgfplotsset{every tick label/.append style={font=\scriptsize}}
    %\pgfplotsset{every axis label/.append style={font=\small}}
    \begin{tikzpicture}
    \begin{axis}[ymin=-6,ymax=6,axis y line=none,xmax=26,xmin=-4,axis lines = middle,x label style={at={(axis description cs:0.88,0.53)},anchor=north},xlabel=Time,height=5cm, width=\linewidth]
    \sdrow[$T_1$][1](1:2:2.25:20.5:21.75)[\phantom{aaaa}21.75][-0.75]
    \sdrow[$T_2$][2.5](0:1:2.25:18.25:19.25)[\phantom{aaaa}19.25][-0.75]
    \srow[$T_3$][4](2:18.25:19.25:20.5)[\phantom{aaaa}20.5][-0.75]

    \sdrow[$T_1$][-5.5](1.25:2.25:2.5:3.25:4.5)[\phantom{aa}4.5][-0.75]
    \srow[$T_2$][-4](0:1:2.25:3.25)[\phantom{aaa}3.25][-0.75]
    \srow[$T_3$][-2.5](4.5:20.75:21.75:23)[\phantom{aa}23][-0.75]

    %\draw (axis cs:10,9) -- (axis cs:10,-10);
    \node[circle] (x) at (axis cs:10.125,4.5) {\large{$x$}};
    \node[circle] (x) at (axis cs:12.625,-2) {\large{$x$}};

    \draw [decorate,decoration = {brace}] (axis cs:-1.5,1) --  (axis cs:-1.5,5);
    \node[circle,rotate=90] (SRPT) at (axis cs:-3,3) {SRPT}; %Average: 20.5
    \draw [decorate,decoration = {brace}] (axis cs:-1.5,-5.5) -- (axis cs:-1.5,-1.5);
    \node[circle,rotate=90] (Ideal) at (axis cs:-3,-3.5) {Ideal}; %Average: 10.25
    \fill[fill=white](axis cs:0,0.5) rectangle (axis cs:-10,-0.5); %This is a real workaround
    
  \end{axis}

    \end{tikzpicture}
    \caption{A set of tasks for which average completion time under SRPT is $2\times$ higher than an ideal schedule. \tikz\fill[fill=green-fill, draw=green-line](0,0) rectangle (1.5ex,1.5ex); represents a running period, \tikz\fill[fill=orange-fill, draw=orange-line](0,0) rectangle (1.5ex,1.5ex); a waiting period, and \tikz\fill[fill=red-fill, draw=red-line](0,0) rectangle (1.5ex,1.5ex); a blocking period. Some states, like the first blocking period in $T_3$, take no time and are not visible. By extending the length marked $x$ of $T_3$, the completion time of SRPT can be up to $3\times$ worse. The first blocking period of $T_2$ must be longer than $T_1$'s running period to force $T_3$ to start running.}
    \label{fig:srpt-act}
\end{subfigure}

\begin{subfigure}{\linewidth}
    \centering
    \begin{tikzpicture}
    \begin{axis}[ymin=-10,ymax=10,axis y line=none,xmax=13,xmin=-2,ticks=none,axis lines = middle,xlabel=Time]
    \srow[$T_1$][1](2.07:2.22:6.98:8.51)[\cmark][-0.5]
    \sdrow[$T_2$][2.5](0.97:1.97:10.49:11.79:11.89)[\xmark][-0.5]
    \sdrow[$T_3$][4](0:0.97:8.55:10.04:10.14)[\xmark][-0.5]
    \sdrow[$T_4$][5.5](1.97:2.07:7.3:8.51:10.04)[\xmark][-0.5]
    \srow[$T_5$][7](2.22:2.32:10.18:11.79)[\xmark][-0.5]

    \srow[$T_1$][-8](0.1:0.24:5.01:6.54)[\cmark][-0.5]
    \srow[$T_2$][-6.5](0.24:1.25:9.76:9.86)[\cmark][-0.5]
    \srow[$T_3$][-5](1.35:2.32:9.9:10)[\cmark][-0.5]
    \srow[$T_4$][-3.5](1.25:1.35:6.58:8.11)[\cmark][-0.5]
    \sdrow[$T_5$][-2](0:0.1:7.97:8.11:9.72)[\cmark][-0.5]

    \draw (axis cs:10,9) -- (axis cs:10,-9);

    \draw [decorate,decoration = {brace}] (axis cs:-1,1) --  (axis cs:-1,8);
    \node[circle,rotate=90] (SRPT) at (axis cs:-1.6,4.5) {SRPT};
    \draw [decorate,decoration = {brace}] (axis cs:-1,-8) -- (axis cs:-1,-1);
    \node[circle,rotate=90] (Ideal) at (axis cs:-1.6,-4.5) {Ideal};
    \fill[fill=white](axis cs:0,0.5) rectangle (axis cs:-3,-0.5);%This is a real workaround
    
  \end{axis}

    \end{tikzpicture}
    \caption{A concrete set of tasks for which it is feasible to finish 5 times more tasks than SRPT. SRPT finishes only $T_1$, while the oracle produces a schedule that finishes all 5 tasks.}
    \label{fig:srpt-deadline}
\end{subfigure}
%Ben: why isn't this spacing done automatically? :(
\vspace{10pt}

\caption{Solver--generated schedules which for which SRPT performs badly.}
\label{fig:srpt-schedules}
\end{figure}

\begin{figure}[!t]
\centering
\begin{subfigure}{0.32\textwidth}
    \centering
    \includegraphics[width=0.9\textwidth]{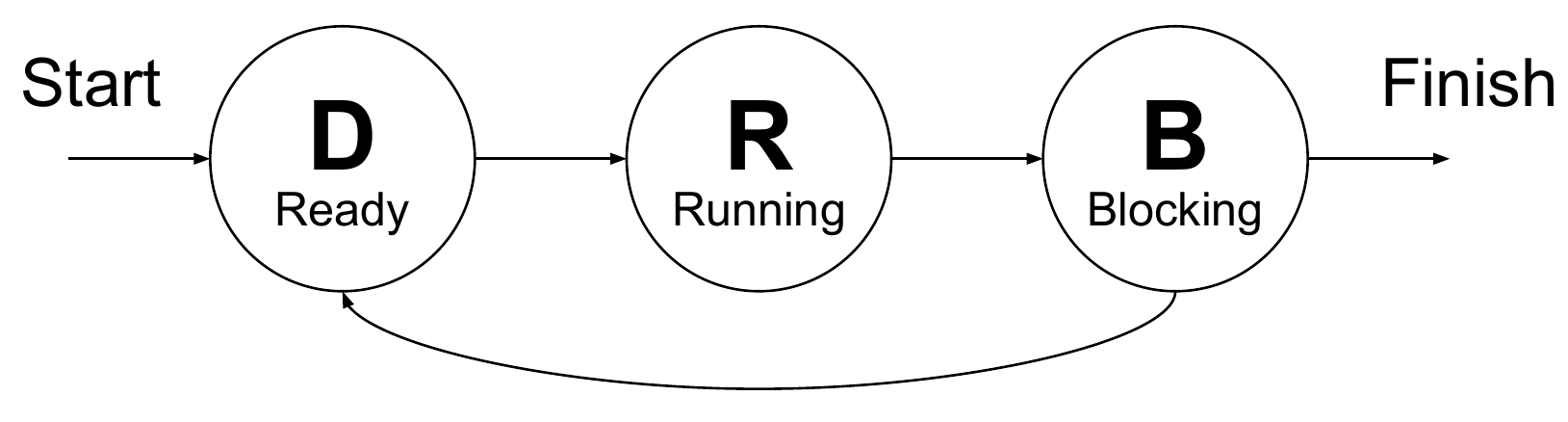}
    \caption{Modeled state machine of a task}
    \label{fig:states_sprt}
\end{subfigure}
\begin{subfigure}{0.45\textwidth}
    \centering
    \includegraphics[width=1\textwidth]{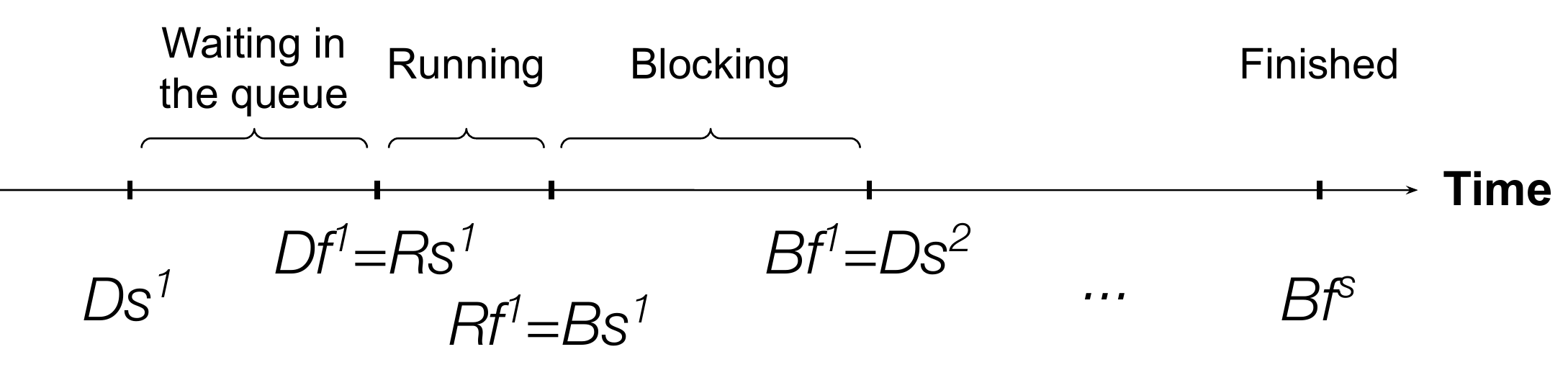}
    \caption{Mapping the state machine to an event-based model in time}
    \label{fig:timeline_srpt}
\end{subfigure}
\caption{Task model, showing the model of a single task}
\label{fig:task_model}
\end{figure}

We leverage \framename to evaluate SRPT scheduling where tasks may block. State is represented in a straightforward manner with task--wise variables for task length, task blocking length, progress made, and the current \textit{stage}: waiting, running or blocking. We number tasks $T_1, \dots, T_n$ and limit the number of events in the model by fixing the number of steps $s$ per task. A step is one ready--running--blocking cycle in Figure~\ref{fig:task_model}; each task has six events per step. We define a task $T_i$ as a tuple of $(L_i,D_i,R_i,B_i)$, where $L_i > 0$ is the total length of the task (i.e., the sum of all the time it spends running). $D,R,B \in (\mathbb{R} \times \mathbb{R})^s$ are sets of pairs that define the start and end of ready, running, and blocking periods, respectively. We denote these with $D = \left\{\left(Ds_i^j, Df_i^j\right)~\mid~1 \leq j \leq s\right\}$, where $Ds_i^j$ is the start of the ready event of task $i$ in step $j$ and $Df_i^j$ is the end of the same ready event. We define $R$ and $B$ similarly.

\begin{figure*}
    \centering
    \vspace{0.25in}
    \includegraphics[width=0.9\textwidth]{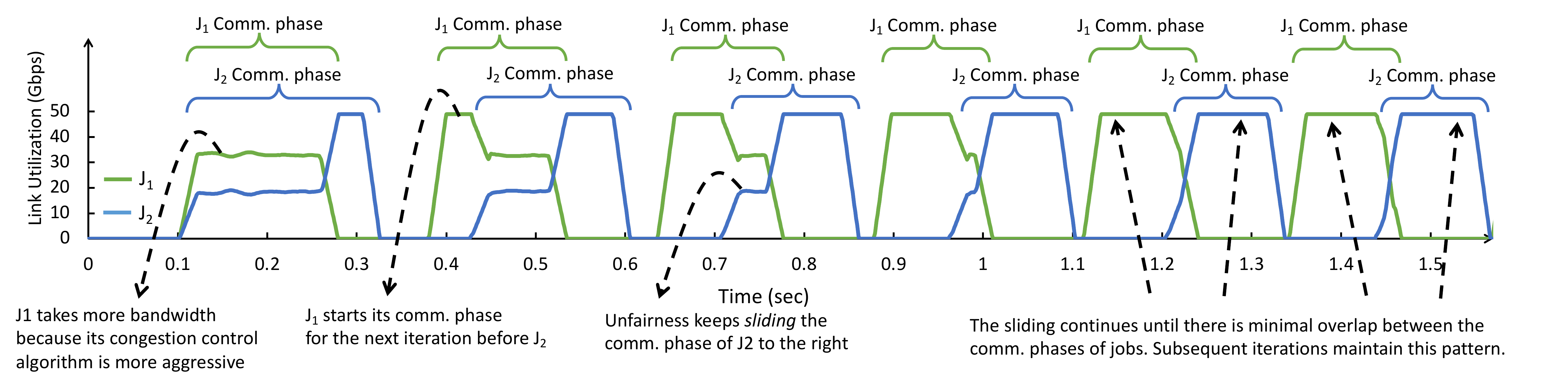}
    \caption{Intuition behind why synchronization occurs. Reproduced from~\cite{ml-net-sync} with permission.}
    \label{fig:rr-intuition}
\end{figure*}

\begin{figure}
\centering
\includegraphics[width=0.6\linewidth]{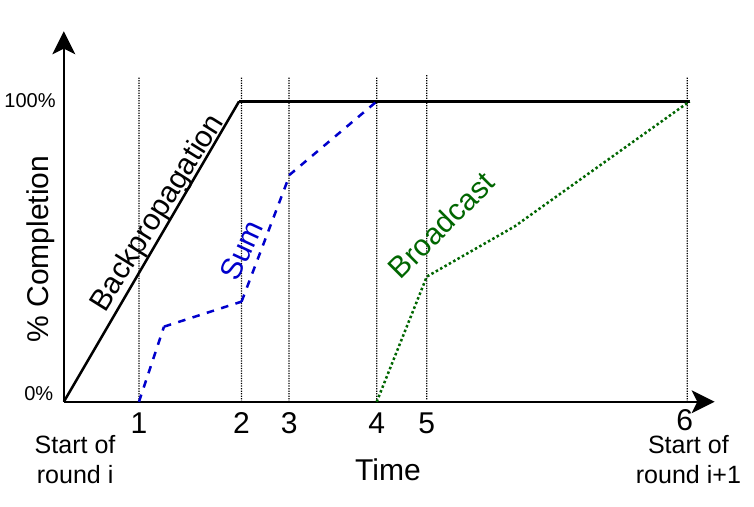}
\caption{Each instance of \statet has a multiple copies of the above variables, one for each server. The figure shows how they could evolve during one round. At {\tt state}$_1$ the server can sends the first $1/n^{th}$ chunk of data as soon as it has computed the gradients for that segment of the network. Till {\tt state}$_2$, the sending rate is lower than the link rate since it is waiting for the previous server to send data. Between {\tt state}$_2$ and {\tt state}$_3$, it is sending at link rate. Between {\tt state}$_3$ and {\tt state}$_4$, it is sharing the link with another job, due to which it sends at a lower rate. Once sum finishes, broadcast can start. Again, the first chunk can be sent without waiting for data from the previous server ({\tt state}$_5$). After this, it sends at less than line rate because it is waiting on data from the previous server, which it forwards as soon as possible.}
\label{fig:ml-training-sync}
\end{figure}

Constraints arising from \framename's \sys ensure the proper timing of all events. At the task level, this means that the sum of the running times per task is equal to its length. Further, tasks must always be in one of the three states, and must proceed from waiting to running to blocking in that order. Formally:
\begin{align*}
&\forall i\quad (Df_i^j = Rs_i^j < Rf_i^j = Bs_i^j \leq Bf_i^j),& 1 \leq j \leq s \\
&\forall i\quad (Bf_i^j =  Ds_i^{j+1})                         ,& 1 \leq j < s 
\end{align*}

There is only one resource, and therefore only one resource queue. The objective function for a task $T$ returns the remaining running time if the task is running or waiting and infinity otherwise, so that currently blocked tasks cannot run. \assign simply changes one task's stage to running. Together, these functions mean that \framename's \algo function is translated into the following constraints, where we let the remaining processing time of task $i$'s step $j$ be
$e_i^j = L_i - \sum_{k=0}^j(Rf_i^k - Rs_i^k)$.

\begin{align*}
&\forall i, l \leq n \hspace{0.1in}  \forall  j,k \leq s \qquad (Rs_i^j < Rs_l^k)  \iff \\
      &\quad\big(e_i^{j-1} \leq e_l^{k-1} \wedge Rf_l^{k-1} \leq Rs_i^j\big) \vee \\
      & \quad\big(e_l^{k-1} \leq e_i^{j-1} \wedge Rf_l^{k-1} \leq Rs_i^j \\
      & \quad\qquad\wedge Bs_l^{k-1} \leq Rs_i^j \wedge Rs_i^j < Bf_l^{k-1}\big) \vee  \\
      & \quad\big(Rf_l^{k-1} > Rs_i^j\big)
\end{align*}

Finally, using \framename, we add constraints to represent the two performance queries of interest. All queries are in the form of two independent schedules $Sched_{SRPT} = \mathcal{T}$ and $Sched_{query} = \mathcal{T}'$, where $Sched_{SRPT}$ follows an SRPT schedule. Both schedules must have the same set of tasks, or $Rf_i^j - Rs_i^j = R'f_i^j - R's_i^j$ and $Bf_i^j - Bs_i^j = B'f_i^j - B's_i^j$ for all $i \leq n$ and $j \leq s$. We specify two queries: 1) comparing the average completion time of tasks under the two schedules, and 2) comparing the number of tasks that finish within a specific deadline.

For average completion time, the query fixes a ratio $q > 0$, and asks whether average completion time in $Sched_{SRPT}$ can be $q$ times more than in $Sched_{query}$:
\begin{equation*}
\left(\sum_{i = 0}^n Bf_i^s\right) = q \left(\sum_{i = 0}^n B'f_i^s\right)
\end{equation*}
For a deadline, the query specifies a time $G$ and compares the number of tasks $a$ and $a'$ finished by time $G$ in each schedule:
\begin{equation*}
\left|\left\{T_i \mid Bf_s^i \leq G\right\}\right|=a~\wedge~\left|\left\{T'_i \mid B'f_s^i \leq G\right\}\right|=a'.
\end{equation*}

Figure~\ref{fig:srpt-schedules} shows the example workloads referenced in Section~\ref{sec:srpt}. Figure~\ref{fig:srpt-act} shows an example set of three tasks for which an ideal scheduler achieves $2\times$ lower average time to completion than SRPT. By extending the running period marked ``x'' in the figure, average completion times can be up to $N_T-\epsilon = 3 - \epsilon$ times worse. Figure~\ref{fig:srpt-deadline} shows a task set for which an ideal scheduler can finish five tasks within a fixed deadline while SRPT finishes only one.

\section{Ring Allreduce Training}\label{app:rr}
\subsection{Model}
We now present the design of our formal model to verify spontaneous synchronization for larger configurations than studied in the reference paper. Here, the heuristic modeled with \framename is the congestion control algorithm (CCA). It is invoked whenever the number of flows sharing a given link changes. It outputs how many bytes each flow gets to transmit in the time between the current and the next step. This way we do not have to model any specific CCA or link type. We only assume that it maintains two key invariants: (1) the link is fully utilized when data is available, and (2) whichever job starts transmitting first gets more bandwidth. This approach captures a broad range of congestion control behaviors~\cite{ml-net-sync}, while keeping the reasoning process simple for both computers and humans.

Figure~\ref{fig:ml-training-sync} shows the \statet variables for a single server at each timestep. These include the percentage of backpropagation, sum, and broadcast finished. It also includes an integer indicating the round number. A round is defined as the index of a batch. Resources are links, and their queues can make up an arbitrary graph, including possibly cycles. To fill in \algo, \objfun simply indicates that all tasks ready to send can be assigned (in this case, more than one task may be identified on each invocation of \algo). \assign allocates bandwidth according to capacity. For the \sys function, \doneCondition is moot---modeling continues until a limit on steps is reached. \nextEvent returns the next time at which a server finishes a computation step or when a flow step finishes sending. Finally, \funUpdate simply updates task variables to reflect progress on the work being done.

Computation and communication are coupled. For example, \textproc{Sum}'s rate of increase is limited not only by the bandwidth allocated by the CCA, but also the progress of computation. For instance, depending on the details of the neural network topology and processor scheduling, it is possible that the server cannot send more than 50\% of the weights until at least 20\% of the backpropagation is complete. \framename guides us on how to over-approximate over these complexities. It says that \sys only needs to interface with \algo. Therefore, we let the solver restrict network transmission in any arbitrary way until \textproc{Backpropagation} reaches 100\%. After that, the server must send at the full capacity allowed by the CCA. In keeping with the theme of this paper, we find that this overapproximation does not sacrifice the provability of our target theorem. We similarly model how \textproc{Backpropagation} depends on \textproc{Sum} sent from the previous round, since it must wait for weights to arrive before starting computation.

\subsection{Spontaneous Synchronization} \label{app:rr-vis}
Figure~\ref{fig:rr-intuition} gives some intuition as to why synchronization occurs spontaneously during Ring Allreduce scheduling.

\end{appendices}